\documentclass{mn}

\usepackage{amsfonts}
\usepackage{amsmath}
\usepackage{graphicx}

\def\gsim{\;\lower4pt\hbox{${\buildrel\displaystyle >\over\sim}$}\;}
\def\lsim{\;\lower4pt\hbox{${\buildrel\displaystyle <\over\sim}$}\;}
\def\grls{\;\lower4pt\hbox{${\buildrel\displaystyle >\over <}$}\;}

\begin{document}

\title[Stationary Structures of two-fluid SIDs]
{Perturbation configurations in a two-fluid system of singular
isothermal disks}
\author[Y.-Q. Lou and Y. Shen]
{Yu-Qing Lou$^{\dagger\ !\ *}$ and Yue Shen$^{\dagger}$ \\
$^{!}$National Astronomical Observatories, Chinese Academy of
Sciences, A20, Datun Road, Beijing, 100012 China\\
$^{\dagger}$Physics Department,The Tsinghua Center for
Astrophysics,Tsinghua
University, Beijing 100084,China\\
$^*$Department of Astronomy and Astrophysics, The University of
Chicago, 5640 S. Ellis Ave., Chicago, IL 60637 USA }
%
\date{Accepted 2003 ... Received 2003 ...;
in original form 2003 ... } \maketitle

\begin{abstract}
We investigate properties of stationary aligned and unaligned
spiral perturbation configurations in a composite system of
gravitationally coupled stellar and gaseous singular isothermal
disks (SIDs) using the two-fluid formalism. Both SIDs are taken to
be razor thin and are in a self-consistent background equilibrium
with power-law surface mass densities
and flat rotation curves. We obtain stationary perturbation
solutions for aligned and unaligned spiral logarithmic
configurations in such a composite SID system and derive
analytically existence criteria for these solutions. In comparison
with the similar problem of a single SID studied by Shu et al.
(2000), there are now two possible sets of solutions owing to an
additional SID. For physically valid solutions, we explore
parameter regimes involving the squared ratio $\beta$ of velocity
dispersions and the ratio $\delta$ of the surface mass densities
of the two SIDs. In terms of transition criteria from axisymmetric
equilibria to aligned secular and spiral dynamical barlike
instabilities, the corresponding ${\cal T}/|{\cal W}|$ ratio of
rotation to potential energies for a composite SID system depends
on $\beta$ and $\delta$, varies in a wide range, and can be
considerably lower than the oft-quoted value $\sim 0.14$. For both
aligned and unaligned cases with azimuthal periodicities
$|m|\geq2$, there exist certain parameter regimes where only one
set of solutions is physically meaningful. For unaligned cases, we
study marginal stabilities for axisymmetric ($m=0$) and
nonaxisymmetric ($m\neq 0$) disturbances. The resulting marginal
instability curves, varying with parameters, are different from
those of a single SID. The case of a composite partial SID system
is also studied to include the gravitational influence of a
dark-matter halo on the system equilibrium. For galactic
applications, our model analysis contains more realistic elements
and offers useful insights for the dynamics of disk galaxies
consisting of stars and gas. Our analytical solutions are valuable
for testing and benchmarking numerical codes. Starting from these
solutions, numerical simulations are powerful to explore nonlinear
dynamics such as large-scale spiral shocks.
\end{abstract}


\section{Introduction}

The theoretical model problem here is one that involves
large-scale stationary density waves in a composite system of
gravitationally coupled stellar and gaseous disks with the two
fluid disks being both idealized as razor-thin singular isothermal
disks (SIDs). For the gravitational effect of a background
axisymmetric dark-matter halo, we also consider a background
composite system of two coupled {\it partial SIDs} (Syer \&
Tremaine 1996; Shu, Laughlin, Lizano \& Galli 2000; Lou 2002). We
search for stationary coplanar perturbation configurations in such
a composite SID system.

Over nearly four decades, there have been numerous theoretical
studies on perturbation and stability properties of a composite
system of stellar and gaseous disks, mostly in galactic contexts.
The local dispersion relation for galactic spiral density waves in
a composite disk system of stars and gas was first derived by Lin
\& Shu (1966, 1968), where the collective behavior of stars was
described by a stellar distribution function while the gaseous
disk was treated as an isothermal fluid. Their pioneer work was
later followed up by Kato (1972), who examined oscillations and
overstability of density waves in a similar formalism. Using the
two-fluid approach, Jog \& Solomon (1984a,b) studied the growth of
local axisymmetric perturbations in gravitationally coupled
stellar and gaseous disks. They found that a composite disk system
can be unstable owing to the gravitational coupling, even though
the stellar and gaseous disks can be separately stable. Bertin \&
Romeo (1988) studied the role of gas on global spiral modes in a
two-fluid model framework. Vandervoort (1991a,b) investigated
effects of interstellar gas on oscillations and stability of
spheroidal galaxies. Romeo (1992) considered the stability of a
two-component disk of finite thickness. The two-fluid formalism
has been adopted into a modal analysis for morphologies of disk
galaxies (Lowe et al. 1994). Elmegreen (1995) and Jog (1996)
simultaneously approached a similar stability problem to derive an
effective $Q_{eff}$ parameter (Safronov 1960; Toomre 1964) for
axisymmetric two-fluid instabilities relevant to a disk galaxy.
Lou \& Fan (1998b) explored basic properties of open and
tight-winding spiral density-wave modes in a composite disk system
using the two-fluid formalism.

Since Mestel (1963), the concept of SIDs has attracted
considerable theoretical interests in various contexts of disk
dynamics in general (Zang 1976; Toomre 1977; Lemos, Kalnajs \&
Lynden-Bell 1991; Syer \& Tremaine 1996; Lynden-Bell \& Lemos
1999; Goodman \& Evans 1999; Shu et al. 2000; Galli et al. 2001).
These studies provide useful information for the research of
star-formation (e.g. Shu et al. 1999 on formation and collapse of
cloud cores in the birth of stars and planetary systems) and
galactic structure communities (e.g., Bertin \& Lin 1996 on the
structure of barred and spiral galaxies and Crane et al. 1993 on
the light cusps seen in the nuclei of galaxies), and deepen our
understanding for the dynamics of self-gravitating configurations
that have a power-law density distribution. In particular, Shu et
al. (2000) derived perturbation solutions and performed a
stability analysis on an isopedically magnetized SID with a flat
rotation curve. The background equilibrium surface mass density
distribution was assumed to bear a power-law radial variation.
They found both stationary aligned and unaligned logarithmic
configurations\footnote{Aligned disturbances involve distorted
streamlines with the maximum and minimum radii at different radial
locations lined up in the azimuth, while for unaligned or spiral
disturbances, distorted streamlines with the maximum and minimum
radii shifted in azimuth at different radial locations (Kalnajs
1973).} and offer physical interpretations for the marginal
instability curves in a single SID. In contrast to their work on a
single SID system, we are mainly interested in the situation of a
disk system composed of a stellar SID and a gaseous SID. As it is
more realistic to consider large-scale dynamics of stellar and
gaseous disks in a disk galaxy, the investigation on such a
composite SID system can reveal more useful information. Motivated
by such a prospect and combining our prior experience of treating
composite disk system (Lou \& Fan 1998b, 2000a, b), we search for
both stationary aligned and unaligned or spiral logarithmic
configurations in a composite SID system and discuss their
stability properties (Safronov 1960; Toomre 1964; Goldreich \&
Lynden-Bell 1965; Ostriker \& Peebles 1973; Elmegreen 1995; Jog
1996; Shu et al. 2000).


Large-scale dynamics of the gas disk is a necessary component in
the overall density-wave scenario. Moreover, most of the
observational diagnostics involve processes of gaseous
interstellar medium (ISM) on various sub-scales. In more realistic
terms, numerical simulation experiments are indispensable for
studying linear and nonlinear dynamical processes. For numerical
code development in particular, the analytical solutions here not
only offer important physical insights but also serve as valuable
tools of benchmarking. Complementarily, numerical simulations
starting from or based upon these stationary perturbation
solutions can lead to insights for equilibrium, stability and
nonlinear processes (e.g., spiral shocks to trigger star formation
activities). Numerical simulations for clusters of galaxies and
numerical simulations for galactic disk dynamics, although on
totally different scales, are similar in several fundamental
aspects, that is, both involve massive dark-matter halo, N-body
gravitational interaction (galaxies as mass points for a cluster
versus stars as mass points for a disk galaxy), and gas dynamics.
The main difference lies in the geometry involved, namely, grossly
spherical geometry for a typical galaxy cluster and a disk
geometry for a spiral galaxy. Therefore, with proper adaptation,
numerical codes designed for a galaxy cluster can be applied to a
disk galaxy or vice versa.


For spiral galaxies, a more realistic model would involve a
magnetized gas disk gravitationally coupled to a stellar disk with
differential rotation in the presence of a massive dark-matter
halo.
For a single magnetized SID (MSID) with a coplanar magnetic field,
we have recently derived solutions for stationary aligned and
logarithmic spiral configurations (Lou 2002) and pointed out slow
MSID configurations can persist in an extended radial range of a
disk with flat rotation curve (Lou \& Fan 2002) as in the case of
NGC 6946 for a spiral pattern of interlaced optical and magnetic
field arms (e.g., Fan \& Lou 1996; Lou \& Fan 1998; Frick et al.
2000). The present hydrodynamic problem is not only interesting by
itself, but also serves as a necessary step for constructing more
realistic stationary configurations in a composite system of a
stellar SID and an MSID.

The paper is structured as follows. In Section 2, we present the
basic two-fluid formalism for the composite system of SIDs coupled
by self-gravity, derive equilibrium properties for both SIDs, and
obtain the linearized equations for small disturbances. Aligned
and unaligned solutions for stationary perturbations in a
composite SID system and their stability properties are studied
and discussed in Section 3. We explore various parameter regimes
for physical solutions and compare them with the results of the
single SID case. The analysis is extended to a composite partial
SID system in Section 4. We summarize computational results and
analysis in Section 5 and discuss potential galactic applications.
Specific details are included in Appendices A$-$D.

\section{Two-fluid SIDs formalism}

For a model containing sufficient physics and for mathematical
simplicity, we adopt the two-fluid formalism for large-scale
stationary aligned or unaligned disturbances in a background
rotational equilibrium with axisymmetry. In dealing with
singularities and resonances, it would be physically more accurate
to adopt the formalism of distribution functions (e.g. Lin \& Shu
1966; Julian \& Toomre 1966; Binney \& Tremaine 1987). For the
purpose of this study, such irregularities and resonances do not
arise and the two-fluid approach will offer useful information for
us to learn. In this section, we provide the basic equations for
the two-fluid system, composed of a stellar disk and a gaseous
disk. Given qualifications and assumptions, equilibrium properties
of the stellar and gaseous SIDs with flat rotation curves (allowed
to be different in a consistent manner) are summarized. We derive
coplanar perturbation equations in both stellar and gaseous SIDs,
respectively.

\subsection{Two sets of coupled fluid equations }

For expediency, the two SIDs located at $z=0$ are treated as
infinitesimally thin, which are sometimes referred to as
razor-thin disks.
For physical variables, we shall use superscripts and/or
subscripts $s$ to denote the stellar disk and superscripts and/or
subscripts $g$ to denote the gaseous disk. The two razor-thin
rotating disks in a composite system are modelled as two fluids
coupled through the mutual gravitational interaction. In the
present formulation of large-scale perturbations, we ignore
nonideal effects such as viscosity, resistivity, and thermal
diffusion etc. Then the two fully nonlinear equation sets for the
stellar and gaseous disks can be written out using cylindrical
coordinates (r,$\varphi,z)$ in the $z=0$ plane. For the stellar
disk, we have
\begin{equation}
\frac{\partial \Sigma^{s}}{\partial t} +\frac{1}{r}\frac{\partial
}{\partial r} (r\Sigma^{s}u^{s})+\frac{1}{r^{2}}\frac{\partial
}{\partial \varphi } (\Sigma^{s}j^{s})=0\ ,
\end{equation}
\begin{equation}
\frac{\partial u^{s}}{\partial t} +u^{s}\frac{\partial
u^{s}}{\partial r} +\frac{j^{s}}{r^{2}}\frac{\partial u^{s}}
{\partial \varphi }-\frac{j^{s2}}{r^{3}}
=-\frac{1}{\Sigma^{s}}\frac{\partial }{\partial
r}(a_{s}^{2}\Sigma^{s}) -\frac{\partial \phi }{\partial r}\ ,
\end{equation}
\begin{equation}
\frac{\partial j^{s}}{\partial t}+u^{s}\frac{\partial
j^{s}}{\partial r} +\frac{j^{s}}{r^{2}}\frac{\partial
j^{s}}{\partial \varphi } =-\frac{1}{\Sigma^{s}}\frac{\partial }
{\partial \varphi }(a_{s}^{2}\Sigma^{s}) -\frac{\partial \phi
}{\partial \varphi }\ .
\end{equation}
In parallel, we have for the gaseous disk
\begin{equation}
\frac{\partial \Sigma^{g}}{\partial t} +\frac{1}{r}\frac{\partial
}{\partial r} (r\Sigma^{g}u^{g})+\frac{1}{r^{2}} \frac{\partial
}{\partial \varphi } (\Sigma^{g}j^{g})=0\ ,
\end{equation}
\begin{equation}
\frac{\partial u^{g}}{\partial t} +u^{g}\frac{\partial
u^{g}}{\partial r} +\frac{j^{g}}{r^{2}}\frac{\partial u^{g}}
{\partial \varphi }-\frac{j^{g2}}{r^{3}}
=-\frac{1}{\Sigma^{g}}\frac{\partial }{\partial
r}(a_{g}^{2}\Sigma^{g}) -\frac{\partial \phi }{\partial r}\ ,
\end{equation}
\begin{equation}
\frac{\partial j^{g}}{\partial t} +u^{g}\frac{\partial
j^{g}}{\partial r} +\frac{j^{g}}{r^{2}}\frac{\partial
j^{g}}{\partial \varphi } =-\frac{1}{\Sigma^{g}}\frac{\partial }
{\partial \varphi }(a_{g}^{2}\Sigma^{g}) -\frac{\partial \phi
}{\partial \varphi }\ .
\end{equation}
The coupling of the two sets of fluid equations is due to the
gravitational potential through Poisson's integral
\begin{eqnarray}
\phi(r,\varphi,t)=
\oint\!d\psi\!\!\int_0^{\infty}\!\!\frac{-G\Sigma (r^{\prime
},\psi ,t)r^{\prime }dr^{\prime }}{\left[ r^{\prime
2}+r^{2}-2rr^{\prime }\cos (\psi -\varphi )\right]^{1/2}}\ ,
\end{eqnarray}
where $\Sigma =\Sigma^{s}+\Sigma^{g}$ is the total surface mass
density. In equations $(1)-(7)$, $\Sigma^{s}$ is the stellar
surface mass density, $u^{s}$ is the radial component of the fluid
velocity, $j^{s}$ is the $z-$component of the specific angular
momentum, and $a_{s}$ is the stellar velocity dispersion (or an
effective ``isothermal sound speed"), $a_{s}^{2}\Sigma^{s}$ stands
for an effective pressure in the polytropic approximation, $\phi $
is the total gravitational potential expressed in terms of
Poisson's integral. For physical variables of the gaseous disk, we
simply replace the superscript $s$ by $g$ systematically. Here, we
assume that the stellar and gaseous disks interact mainly through
the mutual gravitational coupling on large scales (Jog \& Solomon
1984a,b; Bertin \& Romeo 1988; Romeo 1992; Elmegreen 1995; Jog
1996; Lou \& Fan 1998b, 2000a,b).

\subsection{Properties of an axisymmetric equilibrium}

Before a coplanar perturbation analysis, one needs to adopt a
background rotational equilibrium for the composite SID system
consistent with equations $(1)-(7)$. We assume axisymmetric
background SIDs for both stellar and gaseous disks, with the same
form of power-law surface mass densities ($\Sigma\propto r^{-1}$)
yet with different flat rotation curves. The equilibrium
properties of the composite system can then be derived from the
basic equations of Section 2.1. In applications, the divergence
towards $r\rightarrow 0$ may be bypassed by introducing a gradual
transition from disk to bulge or artificial inner cut-outs. In
theoretical analyses, such a $r\rightarrow 0$ divergence poses a
challenge of understanding the SID stability properties (Zang
1976; Toomre 1977; Lynden-Bell \& Lemos 1999; Evans \& Read 1998;
Goodman \& Evans 1999; Shu et al. 2000). In particular, Shu et al.
(2000) have made a systematic investigation on the problem in an
attempt to summarize and clarify the relevant issues and to
resolve the discrepancies between the disparate viewpoints.

Using equations (2) and (5) for the background equilibrium with
$u_0^s=u_0^g=0$, $\Omega_s=j_0^s/r^2$, and $\Omega_g=j_0^g/r^2$,
we readily obtain
\begin{equation}
\begin{split}
\Sigma_0^s=a_s^2(1+D_s^2)/[2\pi Gr(1+\delta)]\ ,\\
\Sigma_0^g=a_g^2(1+D_g^2)\delta/[2\pi Gr(1+\delta)]\ ,
\end{split}
\end{equation}
where $\delta\equiv\Sigma_0^g/\Sigma_0^s$ is the background
surface mass density ratio. Furthermore, one can write
\begin{equation}
\begin{split}
\Omega_s=a_sD_s/r\ ,\\
\Omega_g=a_gD_g/r\ ,
\end{split}
\end{equation}
\begin{equation}
\begin{split}
\kappa_s\equiv\{(2\Omega_s /r)[d(r^2\Omega_s)/dr]\}^{1/2}
=\sqrt{2}\Omega_s\ , \\
\kappa_g\equiv\{(2\Omega_g /r)[d(r^2\Omega_g)/dr]\}^{1/2}
=\sqrt{2}\Omega_g\ ,
\end{split}
\end{equation}
\begin{equation}
a_s^2(D_s^2+1)=a_g^2(D_g^2+1)\ ,
\end{equation}
where $\Omega_s $ and $\Omega_g$ are the mean angular rotation
speeds of the stellar and gaseous SIDs, respectively, $\kappa_s$
and $\kappa_g$ are the corresponding epicyclic frequencies, and
$D_{s}$ and $D_{g}$ are the dimensionless parameters
characterizing the level of rotation for the stellar and gaseous
disks, respectively. The surface mass densities $\Sigma_s$ and
$\Sigma_g$ of both stellar and gaseous SIDs take the power-law
form of $\propto r^{-1}$. Note that condition (11) (also valid for
a composite system of two coupled partial SIDs discussed later) is
derived from the equilibrium radial momentum equations (2) and (5)
using the polytropic approximation for both SIDs. It implies an
intimate relation among the two SID rotation speeds, the stellar
velocity dispersion and the gas sound speed. In this aspect, it is
different from the usual prescription for a composite disk system
(e.g., Jog \& Solomon 1984a, b; Bertin \& Romeo 1988; Elmegreen
1995; Jog 1996; Lou \& Fan 1998b). The two SID rotation speeds
will be different as long as $a_s\neq a_g$. This may induce
streaming instabilities when the difference between the two SID
rotation speeds is sufficiently large.



Here, we introduce two useful parameters for a composite system of
two gravitationally coupled SIDs. The first one is the SID surface
mass density ratio $\delta\equiv\Sigma_{0}^{g}/\Sigma_{0}^{s}$.
For late-type mature spiral galaxies, gas materials are less than
the stellar mass, that is, $\delta<1$. However, for primordial
disk galaxies, the gas materials exceed the stellar mass in
general. Therefore, in our analysis and computations, both cases
of $\delta<1$ and $\delta\geq 1$ are considered. The second
parameter is $\beta\equiv a_s^2/a_g^2$ for the square of the ratio
of the velocity dispersion of stellar disk to the sound speed of
gas disk. Typically, the stellar velocity dispersion is larger
than the sound speed of gas disk. We therefore take $a_s^2>a_g^2$
or $\beta>1$. By ``isothermal", we mean constant $a_s^2$ and
constant $a_g^2$. For actual spiral galaxies, this represents a
gross simplification. In the special situation of $a_s^2=a_g^2$,
it follows from condition (11) that $D_s^2=D_g^2$. Thus, the two
SIDs may be treated as a single SID, because the stellar and gas
disks rotate with the same speed. We will show later that this
$\beta=1$ case gives some familiar results of a single SID which
can be regarded as the limiting case of a two-SID system. In our
analysis, we shall only consider the case\footnote{Background
equilibrium condition (11) guarantees that $D_g^2\geq 0$ as long
as $D_s^2\geq 0$ when $\beta\geq 1$.} of $\beta\geq 1$.

\subsection{Coplanar perturbation equations in SIDs}

We now generally consider small nonaxisymmetric perturbations,
marked along a physical variable with a subscript 1, in both
stellar and gaseous SIDs. For example,
\begin{equation}
\Sigma^{s}=\Sigma {}_{0}^{s}+\Sigma {}_{1}^{s},\qquad
\Sigma^{g}=\Sigma {}_{0}^{g}+\Sigma {}_{1}^{g}\ ,
\end{equation}
\begin{equation}
u^{s}=u_{0}^{s}+u_{1}^{s},\qquad u^{g}=u_{0}^{g}+u_{1}^{g}\ ,
\end{equation}
\begin{equation}
j^{s}=j_{0}^{s}+j_{1}^{s},\qquad j^{g}=j_{0}^{g}+j_{1}^{g}\ ,
\end{equation}
\begin{equation}
\Sigma =\Sigma^{s}+\Sigma^{g} =(\Sigma {}_{0}^{s}+\Sigma
{}_{0}^{g}) +(\Sigma {}_{1}^{s}+\Sigma {}_{1}^{g})\ .
\end{equation}
Substituting expressions $(12)-(15)$ into full equations $(1)-(7)$
and linearizing about the axisymmetric background denoted by a
subscript 0, we derive
\begin{equation}
\frac{\partial \Sigma_{1}^{s}}{\partial
t}+\frac{1}{r}\frac{\partial } {\partial
r}(r\Sigma_{0}^{s}u_{1}^{s}) +\Omega_s \frac{\partial \Sigma
{}_{1}^{s} } {\partial \varphi
}+\frac{\Sigma_{0}^{s}}{r^{2}}\frac{\partial j_{1}^{s}} {\partial
\varphi }=0\ ,
\end{equation}
\begin{equation}
\frac{\partial u_{1}^{s}}{\partial t}+\Omega_s \frac{\partial
u_{1}^{s}} {\partial \varphi }-2\Omega_s \frac{j_{1}^{s}}{r}
=-\frac{\partial }{\partial r}
\bigg(a_{s}^{2}\frac{\Sigma_{1}^{s}}{\Sigma_{0}^{s}}+\phi_{1}\bigg)\
,
\end{equation}
\begin{equation}
\frac{\partial j_{1}^{s}}{\partial t}+r\frac{\kappa_s
^{2}}{2\Omega_s } u_{1}^{s}+\Omega_s \frac{\partial
j_{1}^{s}}{\partial \varphi } =-\frac{\partial }{\partial \varphi
}\bigg(a_{s}^{2}\frac{\Sigma_{1}^{s}} {\Sigma_{0}^{s} }+\phi
_{1}\bigg)
\end{equation}
for the stellar SID, and
\begin{equation}
\frac{\partial \Sigma_{1}^{g}}{\partial
t}+\frac{1}{r}\frac{\partial } {\partial
r}(r\Sigma_{0}^{g}u_{1}^{g}) +\Omega_g \frac{\partial
\Sigma_{1}^{g}}{\partial \varphi }
+\frac{\Sigma_{0}^{g}}{r^{2}}\frac{\partial j_{1}^{g}} {\partial
\varphi }=0\ ,
\end{equation}
\begin{equation}
\frac{\partial u_{1}^{g}}{\partial t}+\Omega_g \frac{\partial
u_{1}^{g}} {\partial \varphi }-2\Omega_g \frac{j_{1}^{g}}{r}
=-\frac{\partial }{\partial r}
\bigg(a_{g}^{2}\frac{\Sigma_{1}^{g}}{\Sigma_{0}^{g}}+\phi_{1}\bigg)\
,
\end{equation}
\begin{equation}
\frac{\partial j_{1}^{g}}{\partial
t}+r\frac{\kappa_g^{2}}{2\Omega_g } u_{1}^{g}+\Omega_g
\frac{\partial j_{1}^{g}}{\partial \varphi } =-\frac{\partial
}{\partial \varphi }\bigg(a_{g}^{2}\frac{\Sigma_{1}^{g}}
{\Sigma_{0}^{g}}+\phi _{1}\bigg)
\end{equation}
for the gaseous SID, with the total gravitational potential
perturbation given by
\begin{eqnarray}
\phi_1(r,\varphi,t)= \oint\!\! d\psi\!\!\int_0^{\infty}\!\!\!
\frac{-G(\Sigma_1^s +\Sigma_1^g)r^{\prime }dr^{\prime
}}{\left[r^{\prime 2} +r^{2}-2rr^{\prime }\cos (\psi -\varphi
)\right]^{1/2}}\ .
\end{eqnarray}
Assuming a Fourier periodic form of $\exp[i(\omega t-m\varphi)]$
for perturbation solutions in general (after taking the real
part), we write for coplanar perturbations in the stellar disk as
\begin{equation}
\begin{split}
\Sigma {}_{1}^{s}=\mu^{s}(r)\exp[i(\omega t-m\varphi)]\ ,\\
u_{1}^{s}=U^{s}(r)\exp[i(\omega t-m\varphi)]\ ,\\
j_{1}^{s}=J^{s}(r)\exp[i(\omega t-m\varphi)]\ ,
\end{split}
\end{equation}
for coplanar perturbations in the gaseous disk as
\begin{equation}
\begin{split}
\Sigma {}_{1}^{g}=\mu ^{g}(r)\exp[i(\omega t-m\varphi)]\ ,\\
u_{1}^{g}=U^{g}(r)\exp[i(\omega t-m\varphi)]\ ,\\
j_{1}^{g}=J^{g}(r)\exp[i(\omega t-m\varphi )]\ ,
\end{split}
\end{equation}
and for the total gravitational potential perturbation as
\begin{equation}
\phi _{1}=V(r)\exp[i(\omega t-m\varphi )]\
\end{equation}
in the SID plane at $z=0$. By substituting expressions $(23)-(25)$
into equations $(16)-(22)$, we derive for the stellar disk
\begin{equation}
i(\omega -m\Omega_s )\mu ^{s}
+\frac{1}{r}\frac{d}{dr}(r\Sigma_{0}^{s}U^{s})-im\Sigma
{}_{0}^{s}\frac{J^{s}}{r^{2}}=0\ ,
\end{equation}
\begin{equation}
i(\omega -m\Omega_s )U^{s}-2\Omega_s \frac{J^{s}}{r}
=-\frac{d}{dr}\bigg(a_{s}^{2}\frac{\mu
^{s}}{\Sigma_{0}^{s}}+V\bigg)\ ,
\end{equation}
\begin{equation}
i(\omega -m\Omega_s )J^{s} +r\frac{\kappa_s ^{2}}{2\Omega_s }U^{s}
=im\bigg(a_{s}^{2} \frac{\mu ^{s}}{\Sigma_{0}^{s}}+V\bigg)\ ,
\end{equation}
for the gaseous disk
\begin{equation}
i(\omega -m\Omega_g )\mu ^{g}
+\frac{1}{r}\frac{d}{dr}(r\Sigma_{0}^{g}U^{g})-im\Sigma
{}_{0}^{g}\frac{J^{g}}{r^{2}}=0\ ,
\end{equation}
\begin{equation}
i(\omega -m\Omega_g )U^{g}-2\Omega_g \frac{J^{g}}{r}
=-\frac{d}{dr}\bigg(a_{g}^{2} \frac{\mu
^{g}}{\Sigma_{0}^{g}}+V\bigg)\ ,
\end{equation}
\begin{equation}
i(\omega -m\Omega_g )J^{g} +r\frac{\kappa_g ^{2}}{2\Omega_g }U^{g}
=im\bigg(a_{g}^{2}\frac{\mu ^{g}}{\Sigma_{0}^{g}}+V\bigg)\ ,
\end{equation}
and for the total gravitational potential perturbation
\begin{equation}
V(r)=\oint d\psi \int _{0}^{\infty } \frac{-G(\mu ^{s}+\mu
^{g})r^{\prime }dr^{\prime }}{\left[ r^{\prime
2}+r^{2}-2rr^{\prime }\cos \psi \right] ^{1/2}}\ .
\end{equation}
We now use equations (27) and (28) to express $U^{s}$ and $J^{s}$
in terms of $\Psi^s\equiv a_{s}^{2}\mu^{s}/\Sigma_{0}^{s}+V$ for
the stellar SID and similarly, use equations (30) and (31) to
express $U^{g}$ and $J^{g}$ in terms of $\Psi^g\equiv
a_{g}^{2}\mu^{g}/\Sigma_{0}^{g}+V$ for the gaseous SID. The
resulting expressions then become
\begin{equation}
U^{s}=\frac{i}{(\omega -m\Omega_s )^{2} -\kappa_s
^{2}}\bigg[-2\Omega_s \frac{m}{r} +(\omega -m\Omega_s
)\frac{d}{dr}\bigg]\Psi^s
\end{equation}
and
\begin{equation}
\frac{J^{s}}{r}=\frac{1}{(\omega -m\Omega_s )^{2}-\kappa_s^{2}}
\bigg[(\omega -m\Omega_s )\frac{m}{r}-\frac{\kappa_s^{2}}
{2\Omega_s }\frac{d}{dr}\bigg]\Psi^s\
\end{equation}
for the stellar SID, and
\begin{equation}
U^{g}=\frac{i}{(\omega -m\Omega_g )^{2}
-\kappa_g^{2}}\bigg[-2\Omega_g \frac{m}{r} +(\omega -m\Omega_g
)\frac{d}{dr}\bigg]\Psi^g
\end{equation}
and
\begin{equation}
\frac{J^{g}}{r}=\frac{1}{(\omega -m\Omega_g )^{2}-\kappa_g^{2}}
\bigg[(\omega -m\Omega_g )\frac{m}{r}-\frac{\kappa_g^{2}}
{2\Omega_g }\frac{d}{dr}\bigg]\Psi^g\
\end{equation}
for the gaseous SID, respectively.

Substitution of expressions (33) and (34) into equation (26) leads
to
\begin{eqnarray}
& &\!\!\!\!\!\!\!\!\! 0=(\omega -m\Omega_s )\mu^{s}
+\frac{1}{r}\frac{d}{dr} \nonumber \\ & & \!\!\!\!\!\!\!\!\!\!\!\!
\times\bigg\{\frac{r\Sigma_{0}^{s}} {(\omega -m\Omega_s
)^{2}-\kappa_s ^{2}}\bigg[-2\Omega_s\frac{m}{r}+(\omega -m\Omega_s
) \frac{d}{dr}\bigg]\Psi^s\bigg\}
\nonumber \\ & & \!\!\!\!\!\!\!\!\!\!\!\!
-\frac{m\Sigma_{0}^{s}}{r[(\omega -m\Omega_s )^{2}-\kappa_s^{2}]}
\bigg[(\omega -m\Omega_s )\frac{m}{r} -\frac{\kappa_s
^{2}}{2\Omega_s }\frac{d}{dr}\bigg]\Psi^s
\end{eqnarray}
for the stellar SID. Similarly, substitution of expressions (35)
and (36) into equation (29) leads to
\begin{eqnarray}
& &\!\!\!\!\!\!\!\!\! 0=(\omega -m\Omega_g )\mu^{g}
+\frac{1}{r}\frac{d}{dr} \nonumber \\ & & \!\!\!\!\!\!\!\!\!\!\!\!
\times\bigg\{\frac{r\Sigma_{0}^{g}} {(\omega -m\Omega_g
)^{2}-\kappa_g ^{2}}\bigg[-2\Omega_g \frac{m}{r}+(\omega
-m\Omega_g ) \frac{d}{dr}\bigg]\Psi^g\bigg\}
\nonumber \\ & & \!\!\!\!\!\!\!\!\!\!\!\!
-\frac{m\Sigma_{0}^{g}}{r[(\omega -m\Omega_g)^{2}-\kappa_g^{2}]}
\bigg[(\omega -m\Omega_g )\frac{m}{r} -\frac{\kappa_g
^{2}}{2\Omega_g }\frac{d}{dr}\bigg]\Psi^g
\end{eqnarray}
for the gaseous SID. Equations (37) and (38) are to be solved with
Poisson's integral (32).

For stationary solutions ($\omega =0$) with zero pattern speed,
equations (37) and (38) above can be cast into the forms of
equations (39) and (40) below by invoking the background
equilibrium conditions $(8)-(11)$. That is,
\begin{eqnarray}
& & m\bigg[-\mu ^{s}
+\frac{1}{D_{s}^{2}(m^{2}-2)}\bigg(\frac{m^{2}}{r}-2\frac{d}{dr}-r
\frac{d^{2}}{dr^{2}}\bigg) \nonumber \\ & &\qquad\qquad\qquad
\times\bigg(r\mu^{s}+\frac{1+D_{s}^{2}}{2\pi G}
\frac{V}{1+\delta}\bigg)\bigg]=0
\end{eqnarray}
for the stellar SID, and
\begin{eqnarray}
& & m\bigg[-\mu ^{g}+\frac{1}{D_{g}^{2}(m^{2}-2)}
\bigg(\frac{m^{2}}{r}-2\frac{d}{dr}-r \frac{d^{2}}{dr^{2}}\bigg)
\nonumber \\ & &\qquad\qquad\qquad
\times\bigg(r\mu^{g}+\frac{1+D_{g}^{2}}{2\pi G} \frac{V\delta
}{1+\delta}\bigg)\bigg]=0
\end{eqnarray}
for the gaseous SID. Again, equations (39) and (40) are to be
solved simultaneously together with Poisson's integral (32).

\section{Aligned and unaligned cases}

\subsection{Aligned perturbation configurations}

Let us now obtain the aligned stationary density wave patterns
from equations (32), (39) and (40). We note in particular that
aligned perturbations relate to purely azimuthal propagations of
density waves (see Section 3.2 of Lou 2002).

\subsubsection{The $|m|=0$ Case: axisymmetric disturbances}

We first examine the $m=0$ case. With $\omega=m=0$, the solution
to equations $(26)-(32)$ takes the forms of $U^s=U^g=0$,
$\mu^s=K_1^s/r$, $\mu^g=K_1^g/r$, $J^s=K_2^sr$, $J^g=K_2^gr$,
$V=K\ln r$, where the ratios of constants $K_1^s/K$, $K_1^g/K$,
$K_2^s/K$, and $K_2^g/K$ are chosen such that equations (27),
(30), (32) can be satisfied. However, such a ``solution" merely
represents a rescaling of one axisymmetric equilibrium to a
neighboring axisymmetric equilibrium (Shu et al. 2000). This
rescaling is allowed by equations (32), (39) and (40) but
uninteresting in the present context. We now turn to cases with
$|m|\geq1$.

\subsubsection{Cases with $|m|\geq 1$:
nonaxisymmetric disturbances}

In power-law disks, we consider aligned perturbations that carry
the same power-law dependence as the equilibrium SID does. By this
assumption, we mean the following exact potential-density relation
\begin{equation}
\begin{split}
\mu_s=\sigma_s/r,\qquad\qquad\\
\mu_g=\sigma_g/r,\qquad\qquad\\
V=-2\pi Gr\mu_s/|m|-2\pi Gr\mu_g/|m|\ ,
\end{split}
\end{equation}
where $\sigma_s$ and $\sigma_g$ are constant coefficients. It can
be verified that Poisson's integral (32) is satisfied (Shu et al.
2000). A direct substitution of equation (41) into equations (39)
and (40) then leads to the following equations:
\begin{equation}
\mu^{s}=\bigg(\frac{m^{2}}{r}-2\frac{d}{dr}
-r\frac{d^{2}}{dr^{2}}\bigg)(H_{1}r\mu^{s} +G_{1}r\mu^{g})\ ,
\end{equation}
\begin{equation}
\mu^{g}=\bigg(\frac{m^{2}}{r}-2\frac{d}{dr}
-r\frac{d^{2}}{dr^{2}}\bigg)(H_{2}r\mu^{g} +G_{2}r\mu ^{s})\ ,
\end{equation}
where coefficients $H_1$, $H_2$, $G_1$ and $G_2$ are defined by
\begin{equation}
\begin{split}
H_1\equiv\frac{1}{D_s^2(m^2-2)}\bigg(1-\frac{D_s^2+1}{|m|}
\frac{1}{1+\delta}\bigg)\ ,\\
H_2\equiv\frac{1}{D_g^2(m^2-2)}\bigg(1-\frac{D_g^2+1}{|m|}
\frac{\delta}{1+\delta}\bigg)\ ,
\end{split}
\end{equation}
\begin{equation}
\begin{split}
G_1\equiv-\frac{D_{s}^{2}+1}{D_{s}^{2}(m^{2}-2)|m|}\frac{1}{1+\delta}\ ,\\
G_2\equiv-\frac{D_{g}^{2}+1}{D_{g}^{2}(m^{2}-2)|m|}\frac{\delta}{1+\delta}\
..
\end{split}
\end{equation}
By substituting the forms of $\mu^s$ and $\mu^g$ given by equation
(41) into equations (42) and (43) above, we readily obtain
\begin{equation}
(1-H_1m^2)\mu^s=G_1m^2\mu^g
\end{equation}
and
\begin{equation}
(1-H_2m^2)\mu^g=G_2m^2\mu^s\ .
\end{equation}
It then follows from equations (46) and (47) that
\begin{equation}
(1-H_1m^2)(1-H_2m^2)=G_1G_2m^4\ .
\end{equation}
In the above procedure, we have assumed $\delta\neq 0$ in order to
eliminate $\mu^s$ and $\mu^g$ from both sides of equation (48).
Otherwise the problem would reduce to that of a single SID.

Using definitions (44) and (45) for the expressions of $H_1$,
$H_2$, $G_1$ and $G_2$, we obtain explicitly
\begin{equation}
\begin{split}
\bigg[1-\frac{m^2}{D_s^2(m^2-2)}
\bigg(1-\frac{D_s^2+1}{|m|}\frac{1}{1+\delta}\bigg)\bigg]\qquad\\
\times\bigg[1-\frac{m^2}{D_g^2(m^2-2)}\bigg(1-\frac{D_g^2+1}{|m|}
\frac{\delta}{1+\delta}\bigg)\bigg]\\
=\frac{D_{s}^{2}+1}{D_{s}^{2}(m^{2}-2)}
\frac{D_{g}^{2}+1}{D_{g}^{2}(m^{2}-2)}\frac{m^2\delta}{(1+\delta)^2}
\end{split}
\end{equation}
which can be further simplified to
\begin{equation}
\begin{split}
\bigg[D_s^2(m^2-2)-m^2\bigg(1- \frac{D_s^2+1}{|m|}
\frac{1}{1+\delta}\bigg)\bigg]\quad\\
\times\bigg[D_g^2(m^2-2) -m^2\bigg(1-\frac{D_g^2+1}{|m|}
\frac{\delta}{1+\delta}\bigg)\bigg]\\
=(D_s^2+1)(D_g^2+1)\frac{m^2\delta}{(1+\delta)^2}\ .
\end{split}
\end{equation}
By condition (11) for the background equilibrium, we have
$D_g^2=\beta(D_s^2+1)-1$ with $\beta\equiv a_s^2/a_g^2$. Equation
(50) can be cast into a quadratic equation in terms of $y\equiv
D_s^2$, namely
\begin{equation}
C_2y^2+C_1y+C_0=0\ ,
\end{equation}
where
$$C_2=\beta(m^2-2)(m^2+|m|-2)\ , \eqno(51a)$$
\begin{equation}
\begin{split}
C_1=\bigg(-2{m}^{2}+4-{\frac{2m\delta}{1+\delta}}-{\frac
{4m}{1+\delta}}+{\frac{2m^3}{1+\delta}}\bigg)\beta
\qquad\qquad \\
\qquad\qquad\hbox{ } +{\frac{2m}{1+\delta}}-2{m}^{4}+6{m}^{2}
-{\frac{2m^3}{1+\delta}}-4\ ,\nonumber \qquad\qquad  (51b)
\end{split}
\end{equation}
\begin{equation}
\begin{split}
C_0=\bigg({\frac{{m}^{3}}{1+\delta}}+2{m}^{2}-{m}^{4}
-{\frac{2m}{1+\delta}}-{\frac{{m}^{3}\delta}{1+\delta}}\bigg)\beta
\qquad\qquad \\
\qquad\qquad\qquad +{\frac{2m}{1+\delta}}-2{m}^{2}
-{\frac{2m^3}{1+\delta}}+2{m}^{4}\ .\nonumber\qquad\qquad (51c)
\end{split}
\end{equation}
The physical meaning is that for aligned nonaxisymmetric
stationary configurations to exist in a composite SID system,
condition (50), or equivalently, condition (51) must be satisfied
for proper values of $D_s^2$ given specified parameters $m$,
$\delta$ and $\beta$. In essence, a purely azimuthal propagation
of density wave in opposite direction relative to the SID rotation
needs to be counterbalanced by disk rotation in order to appear
stationary in an inertial frame of reference. There are two
possible sets of density waves in a composite disk system (e.g.,
Lou \& Fan 1998b). Therefore, there are two possible disk rotation
speeds that may produce stationary perturbation configurations in
general.

Mathematically, one can show that equation (51) always has two
real solutions for $y\equiv D_s^2$, except for the special case of
$m=1$. The two real solutions are not necessarily both physically
valid due to the nonnegative requirement of $D_s^2\geq0$.
Nevertheless, it can be proven later that there always exists at
least one physical solution with $y\equiv D_s^2\geq0$.

For the aligned case of $m=1$, it turns out that $C_2=C_1=C_0=0$
by definitions $(51a)-(51c)$ and
$\mu^g/\mu^s=\Sigma_0^g/\Sigma_0^s$ according to equation (46).
Equation (51) can therefore be satisfied for arbitrary values of
$D_s^2$. This is quite similar to the $m=1$ case of a single SID
studied by Shu et al. (2000). While noting earlier concerns of
spurious artifacts of the analytical method about this result
(Zang 1976; Toomre 1977), Shu et al (2000) expressed a strong
belief that such aligned eccentric displacements are possible
alternative states of equilibria for extended SIDs. Whether such
$m=1$ case represents a trivial translation of the origin of
coordinates, we note for example that contours of surface mass
density (background plus perturbation ones) are given by
$\Sigma=(\Sigma_0r)/r+\sigma\cos{\varphi}/r=$constant according to
equation (41), while contours of surface mass density resulting
from a small shift $a$ along the $x-$axis of the origin from
$x=0$, $y=0$ to $x=a$, $y=0$ are given by
$\Sigma=[(\Sigma_0r)/r](1+a\cos{\varphi}/r)=$constant. It is then
clear that $m=1$ aligned perturbations are not trivial translation
of the origin of coordinates for the present model
problem.\footnote{In analyzing stability properties of Maclaurin
disks (Takahara 1976; Smith 1979), the marginal stationary case of
$m=1$ represents a trivial translation of the origin of
coordinates involving a uniform velocity perturbation as shown in
footnote $14$ on p319 of Binney \& Tremaine (1987). Using the same
analytical criterion here, one can demonstrate that contours of
the surface mass density (see eqns 5-99 and 5-117 of Binney \&
Tremaine 1987) are equivalent to a small shift of the origin of
coordinates.} On this ground, it seems plausible that stationary
aligned eccentric $|m|=1$ configurations are possible alternative
equilibria for a composite SID system, at least mathematically. We
note, however, for a single partial SID (Lou 2002) and for a
composite system of two coupled partial SIDs (discussed later),
such stationary aligned eccentric $|m|=1$ configurations are not
allowed. In other words, stationary aligned eccentric $|m|=1$
configurations might be possible for protostellar disks where
dark-matter halos are not involved, but may not occur for disk
galaxies where massive dark-matter halos are known to exist.

We now study cases of $m\geq2$. As noted earlier, in a real spiral
galaxy, the stellar velocity dispersion $a_s$ is usually larger
than the sound speed $a_g$ of the gas disk\footnote{For example,
one may take $a_g=7\hbox{km s}^{-1}$ and $a_s=30\hbox{km s}^{-1}$
for a typical late-type spiral galaxy.}, that is, $\beta>1$ in our
model. Theoretically, the case of $\beta=1$ means that the stellar
disk and gaseous disk have the same velocity dispersion, and thus
the same rotation parameter $D_s=D_g$ by condition (11). In some
sense, we may then treat the two SIDs as one single SID with
$\Sigma=\Sigma^s+\Sigma^g$. It is expected that the solution
should have something in common with a single SID system (Shu et
al. 2000). We consider below the case of $\beta=1$ for analytical
solutions and for insights of general solution properties.

With $\beta=1$, equations (50) or (51) can be reduced to
\begin{equation}
(|m|-1)[D_s^2(m^2-2)-m^2][(|m|+2)D_s^2-|m|]=0\
\end{equation}
for arbitrary $\delta$ values. Note that expressions $(51a)$,
$(51b)$ and $(51c)$ are independent of $\delta$ when $\beta=1$. We
therefore have two solutions $D_s^2=m^2/(m^2-2)$ and
$D_s^2=|m|/(|m|+2)$ for $|m|\geq2$, and arbitrary values of
$D_s^2$ for $|m|=1$. Note that subsonic rotation
$D_s^2=|m|/(|m|+2)$ for $|m|\geq2$ is simply the result of a
single SID (equation (27) of Shu et al. 2000), and the supersonic
rotation $D_s^2=m^2/(m^2-2)$ for $|m|\geq2$ is a novel feature of
a composite SID system.

For the phase relationship between the two surface mass
perturbations $\mu^s$ and $\mu^g$, we note that equation (46) can
be written in the form of
\begin{equation}
\frac{\mu^g}{\mu^s}=\frac{1-H_1m^2}{G_1m^2}
=-1-\frac{[D_s^2(m^2-2)-m^2](1+\delta)}{|m|(D_s^2+1)}\ .
\end{equation}
For supersonic rotations with $D_s^2=m^2/(m^2-2)$, one simply has
$\mu^g/\mu^s=-1$, while for subsonic rotations with
$D_s^2=|m|/(|m|+2)$, one has $\mu^g/\mu^s=\delta$ (Lou \& Fan
1998b).

We have noted earlier (Lou \& Fan 2002; Lou 2002) that stationary
aligned nonaxisymmetric configurations in an inertial reference
frame correspond to purely azimuthal propagations of density waves
counterbalanced by SID rotation. Physically, $\mu^g=-\mu^s$ simply
means that surface mass density perturbations of stellar and
gaseous SIDs are completely out of phase to reduce the effect of
gravity. This in turn implies a faster azimuthal density wave
speed and thus requires a faster disk rotation (larger $D_s^2$) to
maintain a stationary configuration.
The case of $\mu^g=\delta\mu^s$ means that mass disturbance in
gaseous disk is in phase with the mass disturbance in stellar
disk, and their ratio is the same as the ratio of the background
surface mass densities of the two SIDs. As the effect of
self-gravity is enhanced, the azimuthal density waves speed is
slower and thus a slower disk rotation (smaller $D_s^2$) is needed
to sustain a stationary configuration.

For more realistic situation of $\beta>1$ in general, we readily
obtain two branches of solution for $y\equiv D_s^2$ from quadratic
equation (51) as functions of $\beta$ when $|m|\geq 2$ and
$\delta$ are specified. It is found that both solution branches
(upper branch $y_1$ and lower branch $y_2$) monotonically decrease
for increasing values of $\beta$, and asymptotically approach
different limits when $\beta$ goes to infinity (see three
numerical examples shown in Fig. 1 and Appendix B for the
variation trends).
\begin{figure}
\begin{center}
\includegraphics[scale=0.35]{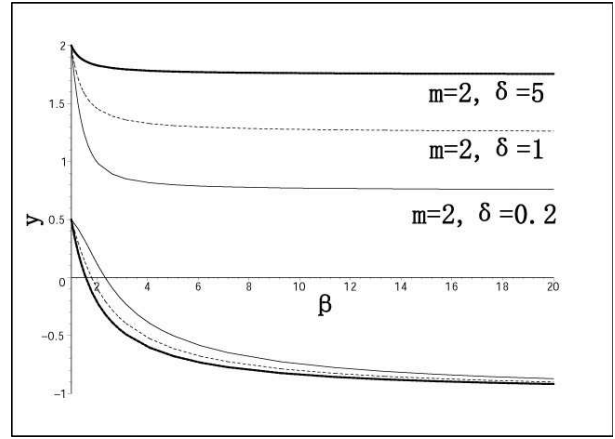}
\caption{Two sets of solution curves $y_1$ and $y_2$ versus
$\beta$ for $|m|=2$ and $\delta=0.2$, $1$, $5$, respectively,
where $\beta$ varies in the interval $[1,\infty)$. The higher
$y_1$ branch is always positive, while the lower $y_2$ branch
becomes negative when $\beta$ exceeds some critical value
$\beta_c$ defined by equation (56) (see also Table 1). Both
solution branches decrease with increasing $\beta$ and approach
different limits as $\beta\rightarrow\infty$. Note also that the
$y_1$ branch approaches the limiting value much faster than the
$y_2$ branch does.}
\end{center}
\end{figure}
It can be shown analytically that the larger solution branch $y_1$
(i.e., the higher branch) satisfies the following inequality
\begin{equation}
\frac{m^2}{m^2-2}-\frac{2m(m+1)}{(m^2-2)(m+2)(1+\delta)} <
y_1<\frac{m^2}{m^2-2}\ ,
\end{equation}
where the lower bound on the left-hand side, which is always
positive, is obtained by taking the limit of
$\beta\rightarrow\infty$ and the upper bound on the right-hand
side is determined by simply taking $\beta=1$. One can also show
analytically that the smaller solution branch $y_2$ (i.e., the
lower branch) satisfies another inequality
\begin{equation}
-1<y_2<|m|/(|m|+2)\ ,
\end{equation}
where the lower bound on the left-hand side is derived by letting
$\beta\rightarrow\infty$ and the upper bound on the right is
obtained by simply taking $\beta=1$. Note that the left-hand side
of $y_1$ in inequality (54) is always greater than the right-hand
side of $y_2$ in inequality (55), so that the two solution
branches will not intersect with each other and $y_1$ remains
always greater than $y_2$. In other words, for a given $|m|$,
while there are two possible aligned solutions at maximum, there
is only one aligned configuration that a composite SID system can
support at a time. However, this does not exclude the possibility
that for different values of $|m|$, more than one aligned
configurations may be sustained in a composite SID system
simultaneously.

By inequalities (54) and (55) and by numerical examples shown in
Figure 1, the lower branches are subsonic while the upper branches
are positive but may vary from supersonic ($\beta\rightarrow 1$)
to subsonic for sufficiently large $\beta$ values when $\delta$ is
small enough (see the case of $m=2$ and $\delta=0.2$). For
late-type spiral galaxies, $\delta$ ranges from 0.05 to 0.1 and
$\beta$ may take values between $\sim 15-20$. Based on the
variation trend displayed in Fig. 1, it is then possible for
late-type disk galaxies to support stationary bar configurations
of the upper branch with subsonic SID rotations. It is also
interesting to infer that for early-type young disk galaxies with
large values of $\delta$, stationary bar configurations of the
upper branch may be sustained by supersonic SID rotations.

By examples of Fig. 1, it also becomes clear that for cases of
$|m|\geq 2$, the lower solution branch $y_2$ of equation (51) may
become negative when $\beta$ is larger than a critical value
$\beta_c$ for given values of $m$ and $\delta$. In other words,
there is only one solution branch for $\beta>\beta_c$, i.e. the
upper $y_1=D_s^2$ branch, which is physically meaningful; being
negative, the lower $y_2=D_s^2$ branch becomes unphysical and
should be discarded.
This critical value $\beta_c$ can be determined analytically in
terms of $\delta$ and $m$ as
\begin{equation}
\displaystyle \beta_c=\frac{2(m+1)}{(m+2)}
\bigg[1+\frac{m}{(m^2+2m)\delta+m^2-2}\bigg]\ .
\end{equation}
For a given value of $m$, the critical value of $\beta_c$
decreases with increasing values of $\delta$.
When $\delta$ approaches infinity, the critical value $\beta_c$
goes to a limiting value
\begin{equation}
\beta_{cLim}=2(m+1)/(m+2)\ .
\end{equation}
In contexts of disk galaxies, this seems to suggest that aligned
barred configurations with $m=2$ of the lower solution branch may
be sustained by subsonic SID rotations only when $\beta < 3/2$
which is rather restrictive.

We now examine the phase relationship between the surface mass
density perturbations $\mu^g$ and $\mu^s$ in general. One can
demonstrate that for the upper $y_1=D_s^2$ branch, the inequality
\begin{equation}
-1<\frac{\mu^g}{\mu^s}<
-\frac{|m|\delta}{(m^2+|m|-2)\delta+m^2-2}\
\end{equation}
holds. It is found that $\mu^g/\mu^s$ ratio decreases with
increasing values of $D_s^2$  (see Appendix C) and therefore
increases with increasing values of $\beta$. The lower bound on
the left-hand side of inequality (58) is determined by taking
$\beta=1$ and the upper bound on the right-hand side of inequality
(58) is obtained by letting $\beta\rightarrow\infty$. For the
lower $y_2=D_s^2$ branch, we have the following inequality
\begin{equation}
\delta < \mu^g/\mu^s < |m|(1+\delta)-1
\end{equation}
for perturbation surface mass density ratio $\mu^g/\mu^s$. As the
mass ratio $\mu^g/\mu^s$ also increases with increasing values of
$\beta$, the lower bound on the left-hand side of inequality (59)
is determined by taking $\beta=1$. However, the upper bound on the
right-hand side of inequality (59) is determined according to the
possible maximum value of $\beta$ that makes $D_s^2$ physically
meaningful, i.e. the critical value $\beta_c$ given by expression
(56). It is now clear that the $D_s^2=y_1$ solution branch is
always characterized by a negative $\mu^g/\mu^s$ and the
$D_s^2=y_2$ solution branch is always characterized by a positive
$\mu^g/\mu^s$. In this regard, the $\beta=1$ case which can be
analyzed thoroughly shows a general solution property.


As observational diagnostics, our analysis suggests two different
types of stationary aligned bar configurations in terms of surface
mass density distributions. For stationary bar configurations of
the upper branch solution, gas bars or young stellar bars should
be out of phase relative to old stellar bars. For stationary bar
configurations of the lower branch solution, gas bars or young
stellar bars should overlap with old stellar bars. Another point
of interest is that for late-type spiral galaxies, $\beta$ may be
as large as $\sim 15-20$. By Fig.1, the lower $y_2$ solution
branches are excluded and the secular barlike instabilities more
likely occur along the upper $y_2$ solution branches.

In general, the two ratios $\beta$ and $\delta$ are independent of
each other. When $\delta$ and $\beta$ are specified, equation (51)
can be solved for two values of $D_s^2$ with different values of
$|m|$. We now take on a specific numerical example below (see Fig.
1). For $|m|=2$ and $\delta=5$, equation (51) becomes
\begin{equation}
8\beta{y}^{2}+(-14-6\beta)y-14\beta+22=0\ ,
\end{equation}
which has two solutions for  $D_s^2=y$, namely,
\begin{equation}
y_1=[7+3\beta+(49-134\beta+121{\beta}^{2})^{1/2}]/(8\beta)
\end{equation}
for $-1<\mu^g/\mu^s<-5/11$, and
\begin{equation}
y_2= [7+3\beta-(49-134\beta+121{\beta}^{2})^{1/2}]/(8\beta)
\end{equation}
for $5<\mu^g/\mu^s<11$. For any value $\beta>1$, $y_1$ given by
equation (61) is always positive. For $1\leq\beta\leq 11/7$, $y_2$
given by equation (62) is nonnegative, while when $\beta>11/7$,
$y_2$ becomes negative and thus unphysical. Besides other
constraints, this may imply that in a disk galaxy where the
stellar velocity dispersion is much greater than the sound speed
of the gaseous disk, the $\mu^g/\mu^s<0$ branch may tend to
manifest. Note that the range of $\mu^g/\mu^s$ is also fully
determined by inequality (58) once $|m|$ and $\delta$ are known.

%
For an numerical example of a late-type disk galaxy, we follow the
similar procedure shown above and take parameters $m=2$,
$\delta=0.05$ and $\beta=4$. Now equation (51) yields two
solutions, namely, $y_1=0.6008$ and $y_2=-0.2972$. Apparently, the
second $y_2$ solution should be discarded since $\beta$ exceeds
the critical value $\beta_c$ given by equation (56), i.e.,
$\beta_c=11/4$. The $y_1$ solution is physically meaningful with
$\mu^g/\mu^s=-0.0822$. In this case, stationary aligned
perturbations in gas surface mass density and stellar surface mass
density are out of phase with each other.

To identify the nature of the aligned solution condition (50) or
(51), we now examine the closely relevant case studied by Shu et
al. (2000) and reemphasize the perspective that stationary aligned
perturbation configurations should be regarded as purely azimuthal
propagation of hydrodynamic density waves (Lou 2002; Lou \& Fan
2002). For this purpose, we write solution condition (50) in a
physically suggestive form of
\begin{eqnarray}
& &\quad[\Omega_s^2(m^2-2)-m^2a_s^2/r^2+2\pi G\Sigma_s^0|m|/r]
\nonumber \\ & &\quad \times[\Omega_g^2(m^2-2)-m^2a_g^2/r^2+2\pi
G\Sigma_g^0|m|/r] \nonumber \\ & &\qquad
=4\pi^2G^2\Sigma_0^s\Sigma_0^gm^2/r^2\ .
\end{eqnarray}
The right-hand side of equation (63) represents the mutual
gravitational coupling between the two SIDs. In the absence of
this coupling, the two factors on the left-hand side would give
rise two separate conditions for stationary aligned perturbation
configurations with $|m|\geq 2$ for stellar and gaseous SIDs,
respectively. For a single SID, be it stellar or gaseous, equation
(63) therefore reduces to the form of
\begin{equation}
m^2\Omega^2=\kappa^2+m^2a^2/r^2-2\pi G\Sigma_0|m|/r \ .
\end{equation}
According to the well-known dispersion relation of density waves
derived under the tight-winding or WKBJ approximation (Lin \& Shu
1964, 1966 or equation (39) of Shu et al. 2000), equation (64) can
be recovered by replacing the radial wavenumber $|k|$ with the
azimuthal wavenumber $|m|/r$ and setting $\omega=0$ in an inertia
frame of reference as noted by Lou (2002) in the study of
stationary MHD perturbation configurations in a single MSID. By
this procedure, it is quite clear that equation (64) describes an
azimuthal propagation of hydrodynamic density waves, and a
stationary pattern in the sidereal frame of reference requires
specific values of $D^2$ for different $|m|$ values (Lou 2002; Lou
\& Fan 2002). In reference to dispersion relation (3.20) of Lou \&
Fan (1998b) with a proper adjustment of two different rotation
speeds of SIDs, it is also transparent that equation (63)
represents a purely azimuthal propagation of density waves in a
composite SID system. The two coupled SIDs rotate in such a way to
render the azimuthal density wave pattern stationary in a sidereal
frame of reference.

In the context of a single SID, Shu et al. (2000) suggested that
stationary condition (64) represents the onset of bifurcations of
axisymmetric SIDs to nonaxisymmetrical SIDs that are more
centrally condensed when the rotation rate $D$ is systematically
increased. This is analogous to classical Maclaurin-Jacobi or
Maclaurin-Dedekind bifurcations from spheroids to ellipsoids
(Chandrasekhar 1969; Tassoul 1978) and involve secular
instabilities that may be induced by viscous dissipations or by
gravitational radiation (Bardeen et al. 1977). Shu et al. (2000)
indicated that the instability mechanisms for aligned and spiral
instabilities are fundamentally different and emphasized that wave
propagation plays no role in aligned perturbations in contrast to
spiral perturbations (see discussions in their subsection 3.1).
Base on our analysis, we would like to clarify here that both
types of perturbations involve propagations of density waves, even
though a radial wave propagation for the spiral case may lead to
different instability mechanisms than those without radial wave
propagation for the aligned case. It is also important to note
that for stationary configurations to appear in a sidereal
reference frame, these density waves travel in opposite sense of
SID rotation relative to the SID system and may strike a balance
between wave propagation and SID rotation.

\begin{table}{\ \ \ \ \ \ \ \ \ \ \ \ \ }
\begin{tabular}{cccc}
\hline $|m|$ & $\delta$ & Critical value $\beta_c$ for\\
&&the lower $y_2$ branch\\
\hline 1 & - & -\\
\hline 2 & 0.2 &2.3333\\
&1&1.8000\\
&5&1.5714\\
&...&...\\
&$\infty$&$1.5000$\\
\hline
3 & 0.2 &2.0800\\
&1&1.8182\\
&5&1.6585\\
&...&...\\
&$\infty$&$1.6000$\\
\hline
\end{tabular}
\caption{Critical value $\beta_c$ of the lower $y_2$ branch
solution for stationary aligned configurations in a composite SID
system, determined by equation (56). See Fig. 1 for three
examples.}
\end{table}

\subsubsection{Secular Barlike Instabilities}

Strictly speaking, we have constructed analytically stationary
aligned nonaxisymmetric configurations in a composite SID system.
Are they stable or do they merely represent transition states
between axisymmetric equilibria and nonaxisymmetric
configurations? This is a challenging question. In the single SID
case, it was suggested (Shu et al. 2000; Galli et al. 2001) that
these solutions signal onsets of bifurcations from an axisymmetric
SID to nonaxisymmetric SIDs (such as eccentric, oval, triangular
distortions corresponding to $m=1,2,3$ etc.). Based on
transmisions and overreflections of leading and/or trailing spiral
density waves across the corotation in a time-dependent problem,
Shu et al. (2000) also made a novel suggestion that the condition
for stationary unaligned or logarithmic spiral configurations in a
SID would determine whether spiral density waves may be
swing-amplified (Goldreich \& Lynden-Bell 1965; Fan \& Lou 1997).
In addition, Shu et al. (2000) examined the parameters of these
stationarity conditions in reference to the disk stability
criterion hypothesized by Ostriker \& Peebles (1973) for the onset
of bar-type instabilities (Miller et al. 1970; Hohl 1971; Kalnajs
1972) and estimated Ostriker-Peebles criterion for aligned secular
and spiral dynamic barlike instabilities in a SID. Judging the
similarity and difference between a single SID and a composite SID
system, we examine stability properties for stationary aligned and
unaligned configurations in the same spirit.

It should be noted that onsets of or transitions to barlike
instabilities may also occur at non-zero pattern speeds such as
those of Maclaurin spheroids to Jacobi ellipsoids (Chandrasekhar
1969) or those of Kalnajs disks to bar configurations etc (Binney
\& Tremaine 1987). The most relevant analogy here is perhaps the
transitions from Maclaurin spheroids to stationary Dedekind
ellipsoids with fixed configurations in space (e.g., Chandrasekhar
1969).

Let us first perform an analysis on aligned secular barlike
instability in a composite SID system. Similar to the single SID
case of Shu et al. (2000), the criterion postulated here is
expressed in terms of the ratio of the kinetic energy of SID
rotation ${\cal T}$ to the absolute value of the gravitational
potential energy ${\cal W}$ (Ostriker \& Peebles 1973; Binney \&
Tremaine 1987). In integral forms, we derive
\begin{equation}
{\cal T}\equiv\int_0^R\frac{1}{2}\Sigma_0^s(r\Omega_s)^22\pi
rdr+\int_0^R\frac{1}{2}\Sigma_0^g(r\Omega_g)^22\pi rdr
\end{equation}
and
\begin{equation}
{\cal W}\equiv-\int_0^Rr(\Sigma_0^s+\Sigma_0^g)
\frac{d\phi_0}{dr}2\pi rdr
\end{equation}
by definitions, where $R$ is a radius that is allowed to go to
infinity. To derive the virial theorem for the composite SID
system from the background equilibrium conditions, we write
equations (2) and (5) in the forms of
\begin{equation}
-\Sigma_0^sr\Omega_s^2 =-\frac{d}{dr}(a_s^2\Sigma_0^s)
-\Sigma_0^s\frac{d\phi_0}{dr}
\end{equation}
and
\begin{equation}
-\Sigma_0^gr\Omega_g^2 =-\frac{d}{dr}(a_g^2\Sigma_0^g)
-\Sigma_0^g\frac{d\phi_0}{dr}\ .
\end{equation}
Adding equations (67) and (68), we obtain
\begin{equation}
(\Sigma_0^s\Omega_s^2+\Sigma_0^g\Omega_g^2)r
=\frac{d}{dr}(a_s^2\Sigma_0^s+a_g^2\Sigma_0^g)
+(\Sigma_0^s+\Sigma_0^g)\frac{d\phi_0}{dr}\ .
\end{equation}
Multiplying equation (69) by $2\pi r^2dr$ and integrating from $0$
to $R$, we arrive at
\begin{equation}
2({\cal T}+{\cal U})+{\cal W}=2\pi R^2
[a_s^2\Sigma_0^s(R)+a_g^2\Sigma_0^g(R)]\ ,
\end{equation}
where
\begin{equation}
{\cal U}\equiv\int_0^R(a_s^2\Sigma_0^s+a_g^2\Sigma_0^g)2\pi rdr
\end{equation}
is the thermal energy contained in the composite SID system.
Equation (70) stands for the virial theorem generalized for a
composite SID system.

Using equilibrium equations $(8)-(11)$, the two integrals (65) and
(66) can be expressed as
\begin{equation}
{\cal T}=a_s^4(D_s^2+1)
\frac{D_s^2+(D_s^2+1-1/\beta)\delta}{2G(1+\delta)}R
\end{equation}
and
\begin{equation}
{\cal W}=-\frac{a_s^4(D_s^2+1)^2}{G}R\ .
\end{equation}
For a composite SID system of infinite radial extent, both
integrals diverge as $R\rightarrow\infty$ but their ratio remains
finite. So the ratio of the kinetic energy of rotation to the
absolute value of the gravitational potential energy is
\begin{equation}
\begin{split}
\frac{{\cal T}}{|{\cal W}|}
=\frac{D_s^2+(D_s^2+1-1/\beta)\delta}{2(1+D_s^2)(1+\delta)}
=\frac{1}{2}-\frac{1+\delta/\beta}{2(D_s^2+1)(1+\delta)}\ .
\end{split}
\end{equation}
Equations $(72)-(74)$ can all be explicitly symmetrized with
respect to the parameters of the two SIDs by using background
condition (11). We here use equation (74) as an example to
illustrate this symmetry between parameters of two SIDs, namely,
\begin{equation}
\begin{split}
\frac{{\cal T}}{|{\cal W}|}
=\frac{a_s^2\Sigma_sD_s^2+a_g^2\Sigma_gD_g^2}
{[a_s^2(1+D_s^2)+a_g^2(1+D_g^2)](\Sigma_s+\Sigma_g)}\ . \nonumber
\end{split}
\end{equation}
Note that the value of ${\cal T}/|{\cal W}|$ falls between $0$ and
$1/2$ as usual (Binney \& Tremaine 1987) and increases with the
increase of $D_s^2$. Therefore, for stationary aligned
configurations in a composite system of two coupled SIDs, the two
possible values of $D_s^2$ correspond to two values of ${\cal
T}/|{\cal W}|$ ratio; the larger and smaller values of $D_s^2$
correspond to larger and smaller values of ${\cal T}/|{\cal W}|$
ratio.


To illustrate the physical significance of the above result, let
us examine a specific case when $|m|=2$, $\delta=0.25$ and
$\beta=2$. Equation (51) now yields two solutions $D_s^2=1.0552$
and $D_s^2=0.0948$ (see Fig. 1). Substitution of these two values
of $D_s^2$ into equation (74) in order gives ${\cal T}/|{\cal
W}|=0.2810$ and ${\cal T}/|{\cal W}|=0.0890$, respectively. Based
on numerical simulation experiments (involving $300-$body
particles) for stability of a rotating disk, Ostriker \& Peebles
(1973) suggested empirically that the approximate condition ${\cal
T}/|{\cal W}|\lsim 0.14\pm 0.02$ is necessary but not sufficient
for stability against bar-type instabilities\footnote{Binney \&
Tremaine (1987) seems to suggest both necessity and sufficiency of
this criterion.}. That is, when ${\cal T}/|{\cal W}|\gsim 0.14\pm
0.02$, a disk system would rapidly evolve into bar-type
configurations (Miller et al. 1970; Hohl 1971; Hunter 1977). By
our numerical example above and the analogy to the single SID case
(Shu et al. 2000), the transition criteria from an axisymmetric
equilibrium to aligned secular barlike instabilities via two
different modes (i.e., upper $y_1$ and lower $y_2$ solutions)
correspond two different values of ${\cal T}/|{\cal W}|$ ratio
(larger and smaller than $\sim 0.14$ respectively).
These variations of ${\cal T}/|{\cal W}|$ ratio appear
considerable but not totally surprising because for Kalnajs disks
(Kalnajs 1972), the relevant criterion is ${\cal T}/{\cal
|W|}<0.1268$ for the stability of bar modes (see also discussions
of Binney \& Tremaine 1987).

There are two possible interpretations for the $y_2$ solution with
a fairly low ${\cal T}/|{\cal W}|=0.0890$. First, if the
correspondence between the ${\cal T}/|{\cal W}|$ ratio and the
onset of secular barlike instabilities holds as suggested by Shu
et al. (2000) for a single SID case and if our extension of this
correspondence to a composite SID system is valid, then our
analysis suggests that the threshold of ${\cal T}/|{\cal W}|$
ratio can be lowered considerably in a composite SID system.
Second, as emphasized by Ostriker \& Peebles (1973) in their note
added in proof, there can be other instabilities that occur for
${\cal T}/|{\cal W}|\lsim 0.14\pm 0.02$. If this is true, then the
suspected correspondence between the ${\cal T}/|{\cal W}|$ ratio
and the onset of secular barlike instabilities might be
coincidental in the single SID case. We are inclined towards the
first interpretation although numerical simulations involving a
composite disk system are deemed necessary to resolve this
important issue. At any rate, a composite SID system tends to be
stabilized by introducing a sufficiently massive dark-matter halo
for both solutions (see our analysis on composite partial SID
system later).

For the same value of $\beta=2$ but with a smaller $\delta$ value,
e.g., $\delta=0.1$, we then have $D_s^2=0.8557$ and
$D_s^2=0.2125$, corresponding to ${\cal T}/|{\cal W}|=0.2428$ and
${\cal T}/|{\cal W}|=0.1064$, respectively. To go further for
$\beta=2$ and $\delta=0.05$, we have $D_s^2=0.7500$ and
$D_s^2=0.2857$, corresponding to ${\cal T}/|{\cal W}|=0.2211$ and
${\cal T}/|{\cal W}|=0.1204$, respectively. By this sequence of
three numerical examples with fixed $\beta=2$ and $m=2$, the
tendency is clear that the ratio ${\cal T}/|{\cal W}|$ for $y_1$
decreases but remains fairly high, while the ratio ${\cal
T}/|{\cal W}|$ for $y_2$ gradually approaches the usual estimate
of $\sim 0.14$ as $\delta$ becomes smaller.

We now complement these numerical examples by an analytical
analysis. As defined by equation (74), the ratio of ${\cal
T}/{\cal |W|}$ involves three parameters $\delta$, $\beta$ and
$D_s^2$ explicitly, and increases with increasing values of
$D_s^2$ and $\beta$. Once $\delta$ and $\beta$ are given, the
marginal value of $D_s^2$ can be determined by equation (51) for
the onset of aligned barlike instabilities with $m=2$. We
investigate below variation trends of ${\cal T}/{\cal |W|}$ versus
$\delta$ when $\beta$ is specified.

Let us first start with $\beta=1.5$. For $m=2$, the critical value
$\beta_c$ as given by equation (56) becomes
\begin{equation}
\begin{split}
\displaystyle \beta_c=\frac{3}{2}
\bigg[1+\frac{1}{4\delta+1}\bigg]\ .\nonumber
\end{split}
\end{equation}
The condition of $\beta\leq\beta_c$ (see the lower branch $y_2$
solutions in Fig 1) requires that for a given $\beta$, $\delta$
must fall within a proper range so as to render the $y_2$ solution
branch physically meaningful. It turns out that for $\beta=1.5$,
$\delta$ can take arbitrary values from $0$ to $\infty$. We then
solve equation (51) for aligned stationary perturbations with
$m=2$ and $\beta=1.5$ to obtain the $y_2=D_s^2$ solution branch as
an explicit function of $\delta$, namely
\begin{equation}
\begin{split}
y_2=\frac{3+4\delta-(1+16\delta+16\delta^2)^{1/2}} {4(1+\delta)}\
.\nonumber
\end{split}
\end{equation}
Substitution of this $y_2$ into equation (74) with $\beta=1.5$
yields an expression for marginal ratio ${\cal T}/{\cal |W|}$ as a
function of $\delta$, namely
\begin{equation}
\begin{split}
\frac{\cal T}{\cal |W|}=
\frac{9+16\delta-3(1+16\delta+16\delta^2)^{1/2}}
{6[7+8\delta-(1+16\delta+16\delta^2)^{1/2}]}\ ,\nonumber
\end{split}
\end{equation}
which has a minimum value of $5/36\approx 0.1389$ at
$\delta=2/3\approx 0.67$ (see the upper curve in Fig. 2).

For a somewhat larger $\beta$ such as $\beta=2$, the value of
$\delta$ is no longer arbitrary in order to obtain nonnegative
$y_2$; it turns out that $0<\delta\leq 0.5$. Repeating the same
procedure for the $\beta=1.5$ case, we then derive another
expression for critical ratio ${\cal T}/{\cal |W|}$ in terms of
$\delta$ as
\begin{equation}
\begin{split}
\frac{\cal T}{\cal |W|}
=\frac{-4-11\delta+3(8\delta+9\delta^2)^{1/2}}
{6[-4-5\delta+(8\delta+9\delta^2)^{1/2}]}\ .\nonumber
\end{split}
\end{equation}
Within the interval of $0<\delta\leq0.5$, this ratio attains a
minimum value of $1/12\approx 0.0833$ at $\delta=0.5$ (see the
middle curve in Fig. 2).

By similar considerations for the case of $\beta=2.5$, the
required range of $\delta$ becomes $0<\delta<1/8$, and the
critical ratio ${\cal T}/{\cal |W|}$ attains the minimum value of
1/30 at $\delta=1/8$ (see the lower curve in Fig. 2).

With a further increase of $\beta$, the required range of $\delta$
diminishes. When $\beta>3$, there is no allowed value of $\delta$
that makes $y_2$ physically meaningful.

\begin{figure}
\begin{center}
\includegraphics[angle=0,scale=0.35]{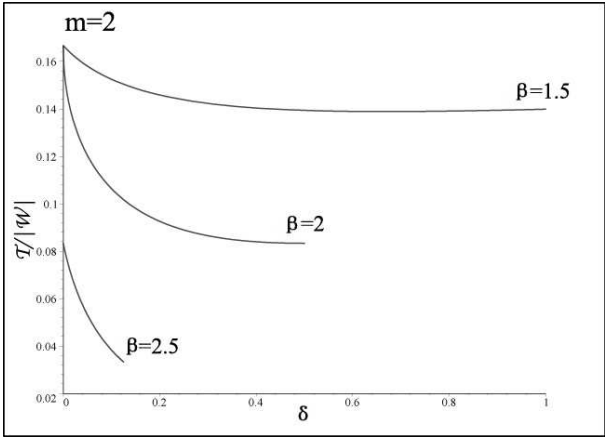}
\caption{Variation trends of the marginal ratio ${\cal T}/{\cal
|W|}$ versus $\delta$ for different values of $\beta=1.5, 2, 2.5$
at a fixed value of $m=2$. Each curve corresponds to an allowed
range of $\delta$ for $y_2=D_s^2\geq0$.}
\end{center}
\end{figure}

These three examples are shown in Figure 2 for the variations of
the marginal ratio ${\cal T}/{\cal |W|}$ versus $\delta$ with
different values of $\beta$ at a fixed value of $m=2$. Each curve
corresponds to an allowed range of $\delta$ for $y_2=D_s^2>0$. Our
analysis of aligned cases already indicate a more complicated
stability properties of a composite SID system than those of a
single SID system. In particular, possible modes of smallest
$D_s^2$ can have different values of $m$ as shown in Figs. 3 and
4.

Starting from a specific example with $\delta=0.2$ and
$\beta=1.5$, the $y_2$ solution branch of aligned equation (51)
yields a curve of $D_s^2$ versus the azimuthal wavenumber $m\geq
2$ as shown in Fig. 3. Qualitatively, this curve differs from the
result of one-SID case, where $D^2=|m|/(|m|+2)$ increases
monotonically with increasing $m\geq2$ (see equation (27) of Shu
et al. 2000 and our equation (52)). Here, $D_s^2$ attains a
maximum at $m=2$ and $3$ and a minimum at $m=7$. In this
particular case, both $m=2$ and $m=3$ give the same value of
$D_s^2$ according to aligned equation (51).

For the same $\delta=0.2$ but a smaller $\beta=1.2$, the curve of
$D_s^2$ for $y_2$ solution branch of equation (51) versus $m\geq2$
is shown in Fig. 4 with the smallest $D_s^2$ at $m=2$. Note the
variation in solution structures in Figs. 3 and 4 for a slight
change of $\beta$ value.
\begin{figure}
\begin{center}
\includegraphics[angle=0,scale=0.29]{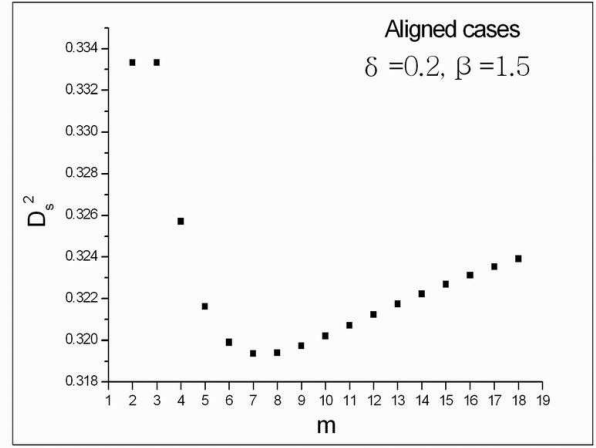}
\caption{The $D_s^2$ solution of $y_2$ branch from aligned
equation (51) versus azimuthal wavenumber $m\geq2$ with parameters
$\delta=0.2$ and $\beta=1.5$. The smallest $D_s^2$ at $m=7$.}
\end{center}
\end{figure}

\begin{figure}
\begin{center}
\includegraphics[angle=0,scale=0.29]{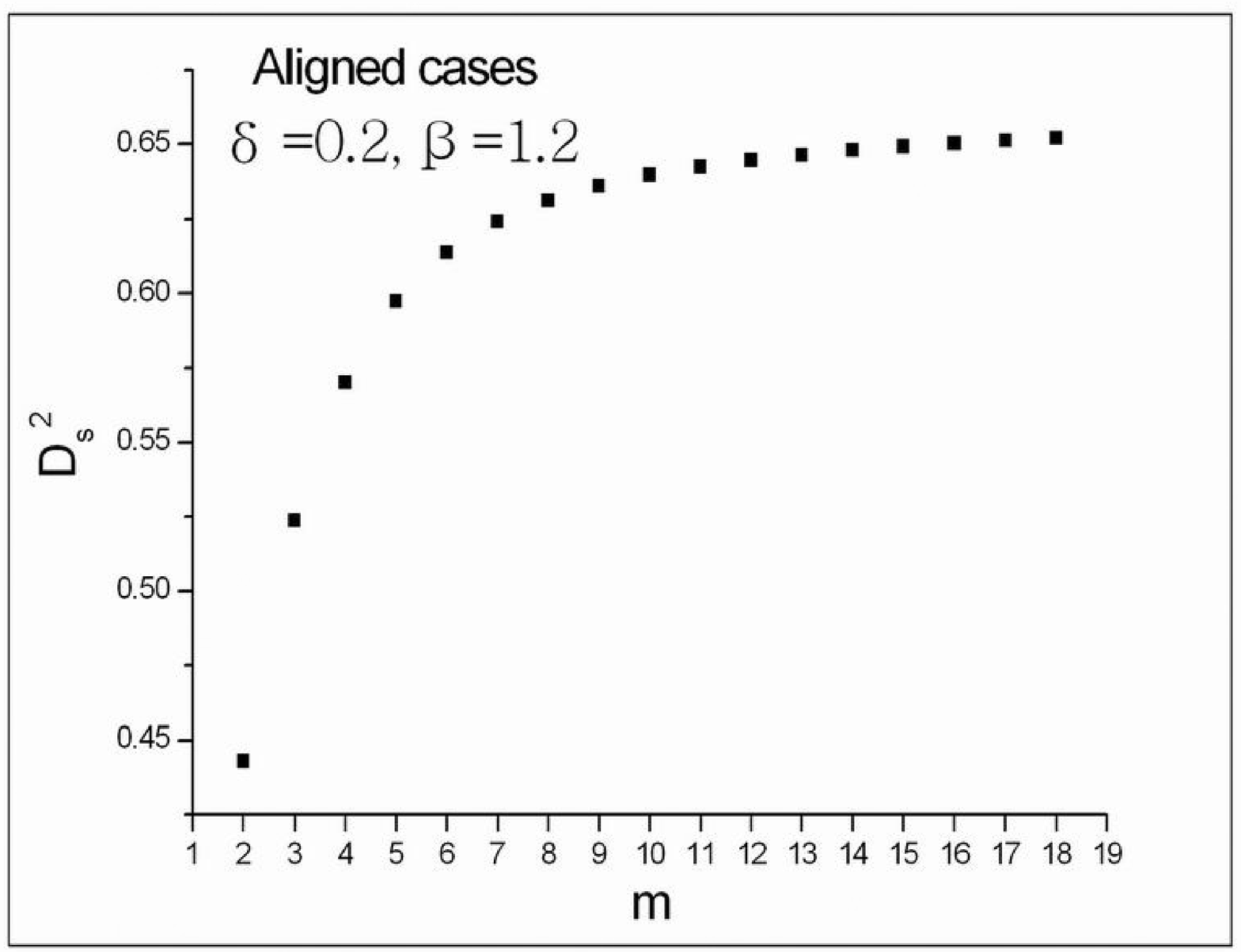}
\caption{The $D_s^2$ solution of $y_2$ branch of from aligned
equation (51) versus azimuthal wavenumber $m\geq2$ with parameters
$\delta=0.2$ and $\beta=1.2$. The smallest $D_s^2$ at $m=2$.}
\end{center}
\end{figure}

%
%

\subsection{Unaligned or spiral disturbances}

For stationary unaligned or spiral disturbances, we take the
following set of exact density-potential relation,
\begin{equation}
\begin{split}
\mu^{s}=\sigma^{s}r^{-3/2}\exp(i\alpha \ln r)\ ,\\
\mu ^{g}=\sigma^{g}r^{-3/2}\exp(i\alpha \ln r)\ ,
\end{split}
\end{equation}
\begin{equation}
V=\upsilon ^{s}r^{-1/2}\exp(i\alpha\ln r) +\upsilon
^{g}r^{-1/2}\exp(i\alpha\ln r)\ ,
\end{equation}
where $\sigma^s$, $\sigma^g$, $\alpha$, $\upsilon^s$, $\upsilon^g$
are constant coefficients and $\upsilon^s$ and $\sigma^s$ and
$\upsilon^g$ and $\sigma^g$ are related by
$$
\upsilon^s=-2\pi G{\cal N}_{m}(\alpha)\sigma^s\ ,
$$
$$
\upsilon^g=-2\pi G{\cal N}_m(\alpha )\sigma^g\ .
$$
Here, ${\cal N}_{m}(\alpha)\equiv K(\alpha,m)$ is the Kalnajs
function (Kalnajs 1971). For spiral perturbations in a composite
SID system, this seems to be a sensible extension of earlier work
relevant to the subject (e.g., Lynden-Bell \& Lemos 1993; Syer \&
Tremaine 1996; Shu et al. 2000; Lou 2002). Note that the radial
scaling parameter $\alpha$ (related to the radial wavenumber) is
naturally taken to be the same in both stellar and gaseous SIDs.


In our analysis and computations, we shall make use of two useful
formula of ${\cal N}_{m}(\alpha)$. One is the recursion relation
in $m$ of ${\cal N}_{m}(\alpha)$ for a fixed $\alpha$ (Kalnajs
1971),
\begin{equation}
{\cal N}_{m+1}(\alpha){\cal N}_{m}(\alpha)
=[(m+1/2)^2+\alpha^2]^{-1}
\end{equation}
and the other is the asymptotic expression of ${\cal
N}_{m}(\alpha)$ (Shu et al. 2000),
\begin{equation}
{\cal N}_{m}(\alpha)\approx(m^2+\alpha^2+1/4)^{-1/2}
\end{equation}
for $m^2+\alpha^2\gg 1$. When accuracy is not that crucial in some
quantitative considerations, asymptotic expression (78) is also
used to compute the $|m|=1$ spiral solutions later.

For $|m|>0$, we proceed to solve equations (39) and (40) which,
with a little algebra, can be cast into the compact forms of
\begin{equation}
\mu ^{s}=\bigg(\frac{m^{2}}{r}-2\frac{d}{dr}
-r\frac{d^{2}}{dr^{2}}\bigg)(H_{1}r\mu^{s} +G_{1}r\mu ^{g})\ ,
\end{equation}
\begin{equation}
\mu^{g}=\bigg(\frac{m^{2}}{r}-2\frac{d}{dr}
-r\frac{d^{2}}{dr^{2}}\bigg) (H_{2}r\mu^g+G_2r\mu^s)\ ,
\end{equation}
where coefficients $H_1$, $H_2$, $G_1$ and $G_2$ are defined by
\begin{equation}
\begin{split}
H_{1}\equiv\frac{1}{D_{s}^{2}(m^{2}-2)}\bigg[1-\frac{(D_{s}^{2}+1)
{\cal N}_{m}(\alpha)}{1+\delta}\bigg]\ ,\\
H_{2}\equiv\frac{1}{D_{g}^{2}(m^{2}-2)}\bigg[1-
\frac{(D_{g}^{2}+1){\cal N}_{m}(\alpha)\delta}{1+\delta}\bigg]\ ,
\end{split}
\end{equation}
\begin{equation}
\begin{split}
G_{1}\equiv -\frac{(D_{s}^{2}+1)}{D_{s}^{2}(m^{2}-2)}
\frac{{\cal N}_{m}(\alpha)}{1+\delta}\ ,\\
G_{2}\equiv -\frac{(D_{g}^{2}+1)}{D_{g}^{2}(m^{2}-2)} \frac{{\cal
N}_{m}(\alpha)\delta}{1+\delta}\ .
\end{split}
\end{equation}
We substitute $\mu^{s}$ and $\mu^{g}$ in the forms of equation
(75) and (76) into equations (79) and (80) to obtain
\begin{equation}
[1-H_{1}(m^{2}+\alpha ^{2}+1/4)]\mu^{s} =G_{1}(m^{2}+\alpha
^{2}+1/4)\mu^{g}\ ,
\end{equation}
\begin{equation}
[1-H_{2}(m^{2}+\alpha ^{2}+1/4)]\mu^{g} =G_{2}(m^{2}+\alpha
^{2}+1/4)\mu^{s}\ .
\end{equation}
Without the gravitational coupling between the stellar and gaseous
SIDs, the result from either disk would be the same as that of Shu
et al. (2000) for a single SID.

By multiplying both sides of equations (83) and (84) and removing
$\mu^s\mu^g$, we obtain
\begin{eqnarray}
[1-H_{1}(m^{2}+\alpha ^{2}+1/4)][1-H_{2}(m^{2}+\alpha^2+1/4)]
\nonumber\\
=G_{1}G_{2}(m^2+\alpha^2+1/4)^2\ .
\end{eqnarray}
By equations (81) and (82), equation (85) becomes
\begin{equation}
\begin{split}
\bigg\{1-\frac{1}{D_{s}^{2}(m^{2}-2)} \bigg[1-\frac{(D_{s}^{2}+1)
{\cal N}_{m}(\alpha )}{1+\delta}\bigg]
\bigg(m^{2}+\alpha^{2}+\frac{1}{4}\bigg)\bigg\}\\
\times\bigg\{1-\frac{1}{D_{g}^{2}(m^{2}-2)}
\bigg[1-\frac{(D_{g}^{2}+1){\cal N}_m(\alpha)\delta}
{1+\delta}\bigg]
\qquad\qquad\\
\times\bigg(m^2+\alpha^2+\frac{1}{4}\bigg)\bigg\}\qquad\quad\\
=\frac{D_{s}^{2}+1}{D_{s}^{2}(m^{2}-2)}
\frac{D_{g}^{2}+1}{D_{g}^{2}(m^{2}-2)} \frac{{\cal
N}_{m}^{2}(\alpha)\delta}{(1+\delta)^2}
\bigg(m^{2}+\alpha^{2}+\frac{1}{4}\bigg)^{2}\ .\qquad
\end{split}
\end{equation}

Equation (86) can be rewritten in a physically more informative
form of
\begin{equation}
\begin{split}
\bigg\{D_s^2(m^2-2)-\bigg[1-\frac{(D_{s}^{2}+1){\cal N}_{m}
(\alpha)}{1+\delta}\bigg]\qquad\qquad\\
\times\bigg(m^{2}+\alpha^{2}+\frac{1}{4}\bigg)\bigg\}\\
\times\bigg\{D_{g}^{2}(m^{2}-2)-\bigg[1-\frac{(D_{g}^{2}+1)
{\cal N}_{m}(\alpha)\delta}{1+\delta}\bigg]\qquad\qquad\\
\times\bigg(m^{2}+\alpha^{2}+\frac{1}{4}\bigg)\bigg\}\\
=(D_{s}^{2}+1)(D_{g}^{2}+1)\frac{{\cal N}_{m}^{2}(\alpha)
\delta}{(1+\delta)^2}\bigg(m^{2}+\alpha^{2}+\frac{1}{4}\bigg)^2\ ,
\end{split}
\end{equation}
which is an exact relation among $D_{s}$, $D_{g}$, $m$, $\alpha$
and $\delta$.

By substituting relation (11), namely $D_g^2=\beta(D_s^2+1)-1$,
into equation (87), we obtain a quadratic equation of $y\equiv
D_s^2$,
\begin{equation}
C_2y^2+C_1y+C_0=0\ ,
\end{equation}
where three coefficients are defined by
\begin{equation}
\begin{split}
C_2=\beta(m^2-2)[m^2-2+(m^2+\alpha^2+1/4) {\cal N}_m(\alpha)]\
,\hbox{ }(88a) \nonumber
\end{split}
\end{equation}
\begin{equation}
\begin{split}
C_1=\bigg[\frac{2+\delta}{1+\delta}(m^2-2)
\bigg(m^2+\alpha^2+\frac{1}{4}\bigg)
{\cal N}_m(\alpha)-\frac{\delta}{1+\delta}\qquad\\
\times\bigg(m^2+\alpha^2+\frac{1}{4}\bigg)^2 {\cal
N}_m(\alpha)-(m^2-2)
\bigg(\alpha^2+\frac{9}{4}\bigg)\bigg]\beta\quad (88b)\\
-\bigg(2m^2+\alpha^2-\frac{7}{4}\bigg)
\bigg[m^2-2+\frac{(m^2+\alpha^2+1/4) {\cal
N}_m(\alpha)}{1+\delta}\bigg]\ ,\hbox{ }\nonumber
\end{split}
\end{equation}
\begin{equation}
\begin{split}
C_0=\bigg(m^2+\alpha^2+\frac{1}{4}\bigg)
\bigg\{(m^2-2)\bigg[\frac{{\cal N}_m(\alpha)}
{1+\delta}-1\bigg]-\frac{\delta}{1+\delta}\quad\\
\times\bigg(m^2+\alpha^2+\frac{1}{4}\bigg) {\cal
N}_m(\alpha)\bigg\}\beta
+\bigg(m^2+\alpha^2+\frac{1}{4}\bigg)\qquad\\
\qquad\qquad \times\bigg(2m^2+\alpha^2-\frac{7}{4}\bigg)
\bigg[1-\frac{{\cal N}_m(\alpha)}{1+\delta}\bigg]\ . \qquad\qquad
(88c) \nonumber
\end{split}
\end{equation}

\subsubsection{The analytical case of $\beta=1$}

Parallel to the analysis for aligned case, we first investigate
the $\beta=1$ case to gain useful insight. When $\beta=1$ and thus
$D_g^2=D_s^2$ by condition (11), equation (88) becomes
\begin{equation}
\begin{split}
[D_s^2(m^2-2)-(m^2+\alpha^2+1/4)]\qquad\qquad\qquad\\
\times\{D_s^2[(m^2-2)+{\cal N}_m(\alpha)
(m^2+\alpha^2+1/4)]\\
+[{\cal N}_m(\alpha)-1](m^2+\alpha^2+1/4)\}=0
\end{split}
\end{equation}
which has two branches of solution
\begin{equation}
D_s^2=y_1=\frac{m^2+\alpha^2+1/4}{m^2-2}
\end{equation}
and
\begin{equation}
D_s^2=y_2=\frac{[1-{\cal N}_m(\alpha)] (m^2+\alpha^2+1/4)}{(m^2-2)
+{\cal N}_m(\alpha)(m^2+\alpha^2+1/4)}\ .
\end{equation}
Meanwhile from equation (83), we have
\begin{eqnarray}
&&\frac{\mu^g}{\mu^s}=\frac{1-H_1(m^2+\alpha^2+1/4)}
{G_1(m^2+\alpha^2+1/4)}
\nonumber\\
&&=-1-\frac{[D_s^2(m^2-2)-(m^2+\alpha^2+1/4)]
(1+\delta)}{(D_s^2+1){\cal N}_m(\alpha)(m^2+\alpha^2+1/4)}\ .
\end{eqnarray}
For $D_s^2=y_1$ in equation (92), one gets $\mu^g/\mu^s=-1$ and
for $D_s^2=y_2$ in equation (92), one gets $\mu^g/\mu^s=\delta$.
These results parallel those of the aligned case. Note that by
solution (90), one has unphysical cases of
$D_s^2<0$ for 
$|m|=1$.
For the solution branch (91) with $|m|=1$, the situation is
somewhat involved. We shall return to this case later at the end
of subsection {\it 3.2.3} around equations $(100)-(104)$.


In general, $\beta>1$ and $\delta$ varies. For different values of
$\delta$ and $\beta$ with $|m|\geq1$, equation (88) or (89) always
yields two solutions of $D_s^2$. For $|m|=1$, one solution is
unphysical.
For $|m|\geq2$, we find that for the two solutions of $D_s^2$, the
ratio of surface mass density perturbations $\mu^g/\mu^s$ is
either positive or negative. This means that for stationary spiral
disturbances in a composite SID system, $\mu^g$ and $\mu^s$ are
either in phase or out of phase. We shall show details later.

Here we have just discussed the special analytic case of $\beta=1$
and point out that the $|m|=1$ case is unique. Our subsequent
computation and analysis for general $\beta$ values are
case-specific from the $|m|=0$ case to the $|m|\geq2$ cases and
then back to the $|m|=1$ case.
\begin{figure}
\begin{center}
\includegraphics[scale=0.35]{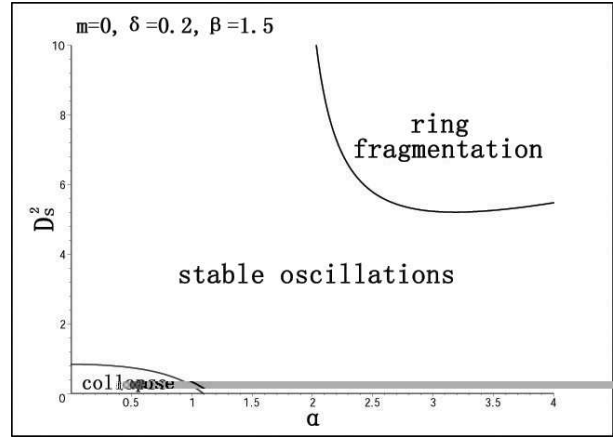}
\caption{The marginal stability curve of $D_s^2$ versus $\alpha$
when $m=0$, $\delta=0.2$, $\beta=1.5$. As $\delta$ is small and
$\beta$ is slightly greater than $1$, the curve does not differ
significantly from the single SID case shown in Fig. 2 of Shu et
al. (2000). Parameter regimes of collapse and ring fragmentation
instabilities as well as stable oscillations are marked.}
\end{center}
\end{figure}

\begin{figure}
\begin{center}
\includegraphics[scale=0.35]{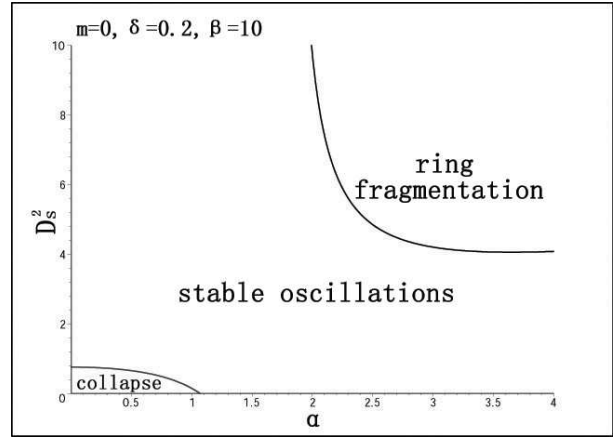}
\caption{The marginal stability curve of $D_s^2$ versus $\alpha$
for $m=0$, $\delta=0.2$, $\beta=10$. While $\delta$ remains small,
a larger $\beta$ lowers the ring fragmentation boundary. It is
easier for the system to become unstable in the form of ring
fragmentation but with a larger $\alpha$. }
\end{center}
\end{figure}

\begin{figure}
\begin{center}
\includegraphics[scale=0.35]{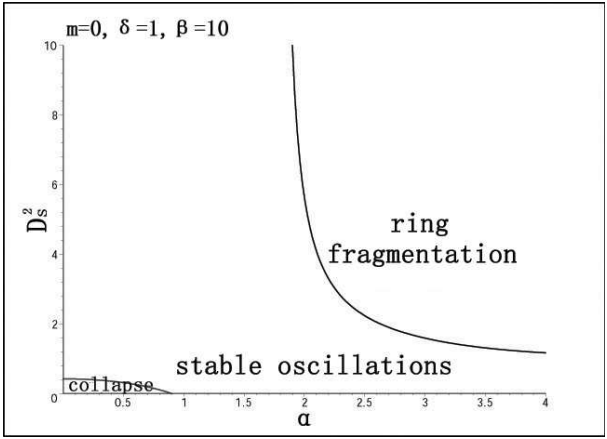}
\caption{The marginal stability curve of $D_s^2$ versus $\alpha$
when $m=0$, $\delta=1$, $\beta=10$. The increase of $\delta$
shrinks the collapse regime at the lower-left corner, while the
increase of both $\delta$ and $\beta$ lowers the marginal
stability curve of ring fragmentation.}
\end{center}
\end{figure}

\begin{figure}
\begin{center}
\includegraphics[scale=1.46]{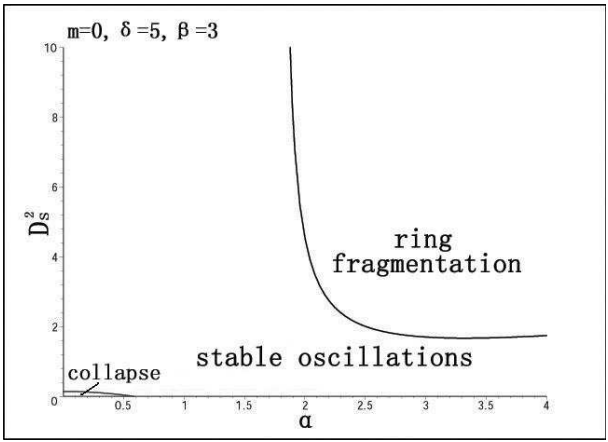}
\caption{The marginal stability curve of $D_s^2$ versus $\alpha$
when $m=0$, $\delta=5$, $\beta=3$. For a moderate $\beta$ value,
the increase of $\delta$ strongly suppresses the collapse regime
while lowers the threshold to ring fragmentation considerably.}
\end{center}
\end{figure}

\begin{figure}
\begin{center}
\includegraphics[scale=0.35]{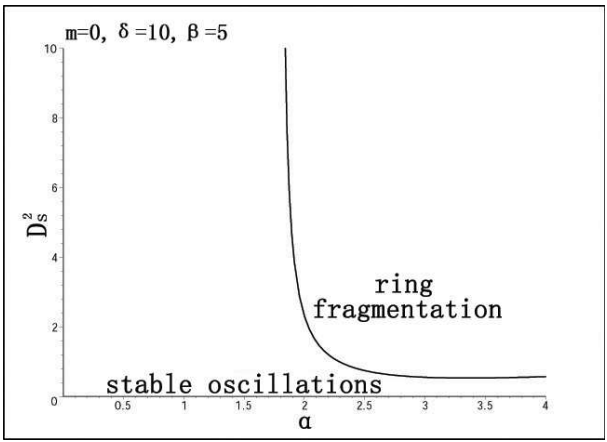}
\caption{The marginal stability curve of $D_s^2$ versus $\alpha$
when $m=0$, $\delta=10$, $\beta=5$. For a sufficiently large
$\delta$, the collapse regime disappears while the danger to ring
fragmentation increases.}
\end{center}
\end{figure}

\begin{figure}
\begin{center}
\includegraphics[scale=0.35]{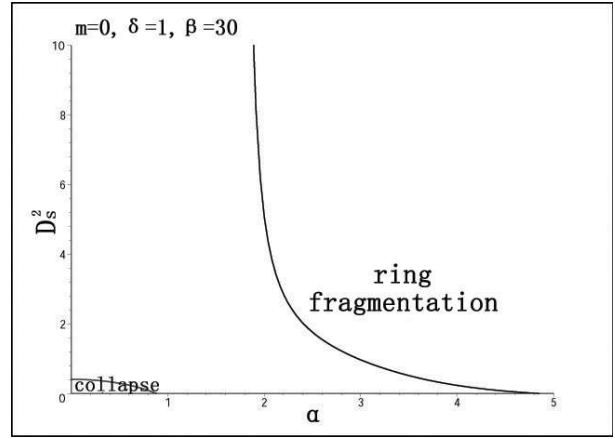}
\caption{The marginal stability curve of $D_s^2$ versus $\alpha$
when $m=0$, $\delta=1$, $\beta=30$. The ring fragmentation curve
crosses the $\alpha$ axis (and becomes negative) such that ring
fragmentation occurs for all $D_s^2$ when $\alpha$ is sufficiently
large. A much reduced collapse regime still exists.}
\end{center}
\end{figure}


\subsubsection{Marginal stability of axisymmetric disturbances}

While equation (88) or (89) is derived for cases of $|m|>0$, it
can also be used to describe the marginal stability for the
special case of $|m|=0$, that is, stationary axisymmetric
disturbances. Equation (89) with $|m|=0$ yields two branches of
solution constrained by the physical requirement\footnote{Not
shown here is a second solution branch of $D_s^2$ which is
negative and thus unphysical (see Figure $D1$ in Appendix D for an
example corresponding to the same parameters of Figure 2 and also
see equation (90) for $\beta=1$ with $m=0$).} of $D_s^2\geq 0$.
For the positive portions of solution $D_s^2$ with not-so-extreme
parameters, we find the basic $D_s^2$ versus $\alpha$ profile is
qualitatively similar to Figure 2 of Shu et al. (2000) as
expected. When $D_s^2$ falls in the range between the maximum of
the collapse regime and the minimum of the ring fragmentation
curve, the composite SID system can support stable oscillations as
indicated in Figs $5-9$. In the limit of $\alpha\rightarrow 0$,
stable ``breathing modes" can be sustained (Lemos et al. 1991).
When $\delta\equiv\Sigma_0^g/\Sigma_0^s$ is small, which means the
gas mass is small compared to the stellar mass, the gas influence
on the stellar disk is weak and the marginal stability will be
slightly modified compared to that of a single SID situation. As
$\delta$ increases, the role of gas disk becomes dynamically more
important. For young disk galaxies during their early epochs of
formation and evolution, the gas materials can be more abundant
than or comparable to stellar mass, the axisymmetric marginal
stability should be quite different from the single SID situation.

As $\delta$ and/or $\beta$ vary, the marginal stability curves
change accordingly as shown in Figures $5-10$. For small $\delta$
and $\beta\approx 1$, the shape of marginal stability curve is
similar to that of a single stellar SID (Shu et al. 2000). The
curve contains two separate branches in the $D_s^2-\alpha$ plane.
The first collapse branch starts at the vertical $D_s^2>0$ axis
with $\alpha=0$ and goes down to the horizontal $\alpha>0$ axis
with $D_s^2=0$. The second ring fragmentation branch has a
vertical asymptote, where the value of $D_s^2$ approaches infinity
at a finite $\alpha$. The curve descends with increasing $\alpha$
to a minimum and the ascends to infinity as
$\alpha\rightarrow\infty$. There are no other positive values for
$D_s^2$ that can be found between the two branches just described.
When $\delta$ is fixed and $\beta$ takes larger values, which
means the stellar velocity dispersion further exceeds the sound
speed of the gas SID, both branches descend, but this trend of
variation is more conspicuous for the ring fragmentation branch.

For $\delta=0.2$, the increase of $\beta$ significantly lowers the
ring fragmentation curve. In comparison, the collapse branch
changes slightly with increasing $\beta$ when $\delta$ is small.
This is shown in Figs 5 and 6. As $\delta$ increases for fixed
$\beta$ value, the collapse branch shrinks with the surrounded
area reduced (see Figs. 6 and 7). The collapse regime continues to
shrink with increasing $\delta$. When $\delta$ becomes
sufficiently large, this branch will completely disappear from the
$D_s^2-\alpha$ diagram (see Figs. 8 and 9), leaving only the ring
fragmentation branch (Fig. 9). Meanwhile for the ring
fragmentation branch, the trend is clear that as $\delta$
increases, the minimum of the marginal stability curve decreases
at larger values of $\alpha$. Eventually, the curve crosses the
$\alpha$ axis and becomes negative when $\delta$ and $\beta$ take
proper values (Fig 10). We conclude that as $\delta$ or $\beta$
increases, both marginal stability curves move downward (towards
smaller $D_s^2$).

It is interesting to note that the location $\alpha$ of the
vertical asymptote for the ring fragmentation branch remains fixed
independent of $\delta$ and $\beta$. By equation (88), $D_s^2$
diverges when $C_2$ defined by equation $(88a)$ vanishes, or
equivalently\footnote{This is also true when magnetic field is
involved in a SID (Shu et al. 2000; Lou 2002; Lou \& Fan 2002).},
${\cal N}_0(\alpha)(\alpha^2+1/4)=2$. By recursion relation (77)
and the asymptotic expression (78), we may choose approximately
\begin{eqnarray}
&&{\cal N}_0(\alpha)=(9/4+\alpha^2)/(1/4+\alpha^2){\cal
N}_2(\alpha)\nonumber\\
&&\qquad\quad
=(9/4+\alpha^2)/[(1/4+\alpha^2)(17/4+\alpha^2)^{1/2}]
\end{eqnarray}
to estimate an $\alpha$ value of $1.7928$ for $C_2=0$ in equation
(88) (Shu et al. 2000; Lou 2002; Lou \& Fan 2002).

We now provide physical interpretations for the two curves of
marginal stability. The parameter $\alpha$ is a measure of radial
wavenumber. Small $\alpha$ corresponds to larger radial
perturbation scale. Therefore the composite SID system is
vulnerable to Jeans' instability when $\alpha$ is sufficiently
small. Because of the angular momentum conservation, a SID
rotation tends to work against gravitational collapse. Therefore,
it requires a smaller $\alpha$ to induce Jeans' collapse in the
presence of disk rotation. When disk rotation is sufficiently fast
($D_s^2$ larger than a critical value at $\alpha=0$), Jeans'
collapse can be completely prevented.

On the other hand, the composite SID system should be also
vulnerable to axisymmetric Toomre-type instability.\footnote{The
original criterion was derived for local axisymmetric
perturbations in a rotating disk (Safronov 1960; Toomre 1964).} In
various galactic contexts, it is of considerable interest to
search for an effective $Q$ parameter for a composite disk system
(Jog \& Solomon 1984a,b; Bertin \& Romeo 1988; Kennicutt 1989;
Romeo 1992; Wang \& Silk 1994; Elmegreen 1995; Jog 1996; Lou \&
Fan 1998b). Shu et al. (2000) noticed that the minimum of the ring
fragmentation branch is effectively related to the Toomre $Q$
parameter (Toomre 1964) in a very close manner. In a magnetized
SID with a coplanar magnetic field, Lou (2002) and Lou \& Fan
(2002) noticed that that the minimum of the MHD ring fragmentation
branch is effectively related to the generalized MHD $Q_M$
parameter (Lou \& Fan 1998a) in a very close manner. Physically,
we therefore expect that the minimum of the ring fragmentation
branch should be intimately related to an effective $Q$ parameter
in a composite disk system (Elmegreen 1995; Jog 1996; Lou \& Fan
1998b). When $D_s^2$ exceeds the minimum of the ring fragmentation
branch, the composite system of two coupled SIDs becomes unstable
to ringlike fragmentation for the range of radial wavenumbers
$\alpha$ between the two points of intersection of the horizontal
line and the the marginal stability curve.




\begin{figure}
\begin{center}
\includegraphics[scale=0.35]{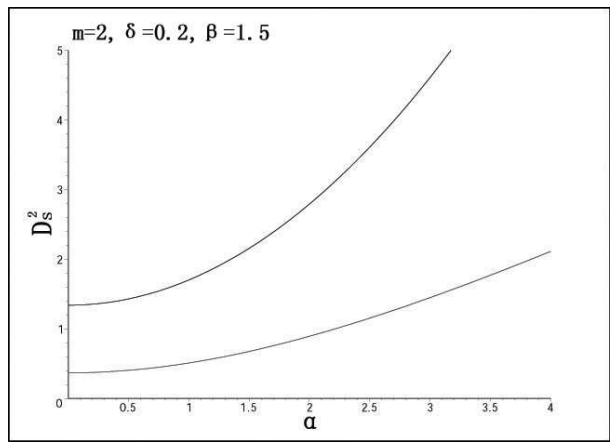}
\caption{Two solutions of stationary spiral configurations for
$D_s^2$ versus radial wavenumber $\alpha$ when $|m|=2$,
$\delta=0.2$, $\beta=1.5$ as derived from equation (88). For small
$\delta$ and $\beta$ slightly greater than $1$, the lower branch
is almost the same as Fig. 3 of Shu et al. (2000). The upper
branch is novel due to a composite disk system. Both curves
increase with increasing $\alpha$.}
\end{center}
\end{figure}

\begin{figure}
\begin{center}
\includegraphics[scale=0.35]{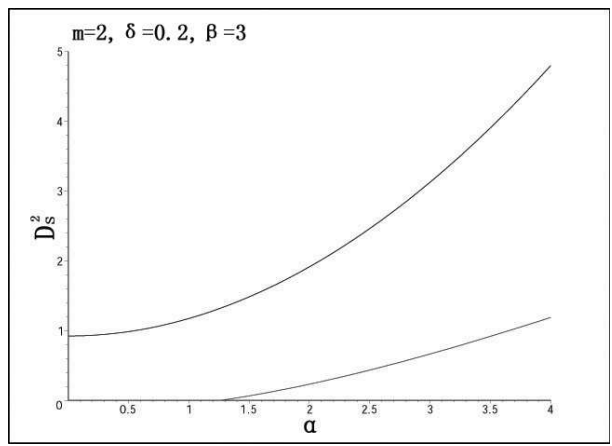}
\caption{Two solutions of stationary spiral configurations for
$D_s^2$ versus radial wavenumber $\alpha$ when $|m|=2$,
$\delta=0.2$, $\beta=3$. As $\beta$ increases, both branches move
downward. It is possible for the lower branch go across the
horizontal $\alpha$ axis, while the upper branch will always
remain positive. }
\end{center}
\end{figure}

\begin{figure}
\begin{center}
\includegraphics[scale=0.35]{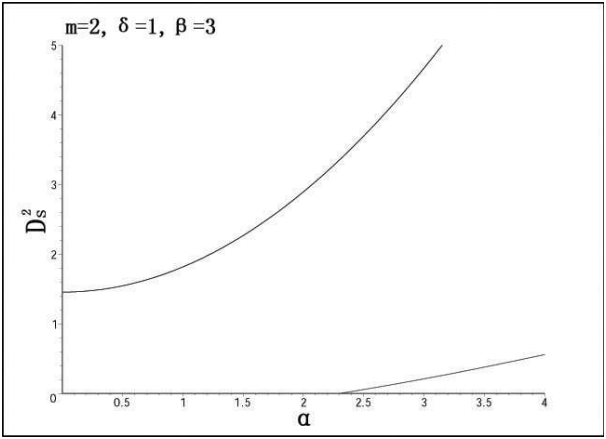}
\caption{Two solutions of stationary spiral configurations for
$D_s^2$ versus radial wavenumber $\alpha$ when $|m|=2$,
$\delta=1$, $\beta=3$. By increasing $\delta$ at fixed $\beta$
(see Fig. 12), the lower branch moves downward while the upper
branch moves upward. }
\end{center}
\end{figure}

\subsubsection{Stationary logarithmic spiral configurations}

Following the same procedure of obtaining solutions for
axisymmetric disturbances, we derive solution curves for cases of
$|m|=1,2,3...$. As expected, for each set of $\{ m$, $\delta$,
$\beta \}$, there are two solution curves for $D_s^2$ versus
$\alpha$ in a composite SID system according to equation (88) or
(89). In some cases, part of the lower solution may become
negative which is unphysical. Figures $11-13$ show a set of
numerical examples with $|m|=2$.


In the case of $|m|=1$, only one branch of $D_s^2$ is physically
meaningful. We will discuss more on this problem later. Let us
first examine cases of $|m|\geq2$.

For given values of $\delta$ and $\beta$, there are two solution
curves in general for $D_s^2$ versus $\alpha$ when $|m|\geq2$. The
lower branch may cross the horizontal $\alpha$ axis when $\delta$
and/or $\beta$ are sufficiently large. In contrast, the upper
branch will never become negative.
We note that variations of $\beta$ and $\delta$ play distinctly
different roles. The increase of $\delta$ mainly moves the upper
branch upward and the lower branch downward, whereas the increase
of $\beta$ tends to move both branches downward. For very small
$\delta$ values, different $\beta$ values may lead to quite
different solution configurations (see Figs. $11-13$).

In order to fully understand solution behaviors (Figs. $11-13$),
we perform analytic analysis on properties of stationary spiral
configurations as in the aligned cases of $|m|\geq2$. Firstly,
through extensive numerical experiments, one can show that
$y\equiv D_s^2$ increases with wavenumber $\alpha$ for both upper
and lower branches. Secondly, once $|m|$ and $\delta$ are known
with fixed $\alpha$, we would like to know how $y\equiv D_s^2$ and
$\mu^g/\mu^s$ vary with $\beta$. It turns out that for both
solution branches, $y\equiv D_s^2$ monotonically decrease with
increasing $\beta$ and $\mu^g/\mu^s$ decreases with increasing
$D_s^2$. It is then clear that $\mu^g/\mu^s$ goes with $\beta$ in
the same trend. In more informative forms, we derive two useful
inequalities for $y_1$ and $y_2$ below. The first one for $y_1$ is
thus
\begin{equation}
\begin{split}
(m^2+\alpha^2+1/4)/(m^2-2)
\qquad\qquad\qquad\qquad\qquad\quad\\
\times\bigg\{1-\frac{[m^2-2+(m^2+\alpha^2+1/4)] {\cal
N}_m(\alpha)}{[m^2-2+(m^2+\alpha^2+1/4)
{\cal N}_m(\alpha)](1+\delta)}\bigg\}\\
<y_1<(m^2+\alpha^2+1/4)/(m^2-2)
\end{split}
\end{equation}
with the lower bound on the left-hand side determined by taking
$\beta\rightarrow\infty$ and the upper bound on the right-hand
side obtained by taking $\beta=1$; and the second one for $y_2$ is
\begin{equation}
-1<y_2<\frac{[1-{\cal N}_m(\alpha)](m^2+\alpha^2+1/4)}{m^2-2+
{\cal N}_m(\alpha)(m^2+\alpha^2+1/4)}
\end{equation}
with the lower bound on the left-hand side determined by taking
$\beta\rightarrow\infty$ and the upper bound on the right-hand
side obtained by taking $\beta=1$. One can readily show that the
lower bound of $y_1$ is positive definite as ${\cal
N}_m(\alpha)<1$ for $|m|\geq 2$. Therefore, the $D_s^2\equiv y_1$
branch is always physically valid given values of $\delta$ and
$\beta$ with $|m|\geq2$. Moreover, as in the aligned cases, the
left-hand side of inequality (94) for $y_1$ remains always larger
than the right-hand side of inequality (95) for $y_2$ such that
the two solution branches $y_1$ and $y_2$ will never intersect
with each other because $y_1>y_2$ always. Meanwhile, there exists
also a critical value $\beta_c$ for $y_2\equiv D_s^2$ branch at a
given wavenumber $\alpha$ such that $y_2\equiv D_s^2$ becomes
zero. In analytic form, this critical value $\beta_c$ is given by
\begin{equation}
\begin{split}
\beta_c=\frac{m^2-2+(m^2+\alpha^2+1/4)}{m^2-2+ {\cal
N}_m(\alpha)(m^2+\alpha^2+1/4)} \bigg\{1+
\qquad\qquad\qquad\\
\frac{{\cal N}_m(\alpha)[1 -{\cal N}_m(\alpha)](m^2+\alpha^2+1/4)}
{{\cal N}_m(\alpha)(m^2+\alpha^2+1/4)\delta+(m^2-2) [1+\delta
-{\cal N}_m(\alpha)]}\bigg\}\ .\quad
\end{split}
\end{equation}
Since ${\cal N}_m(\alpha)<1$ for $|m|\geq2$, the critical value
$\beta_c$ remains always greater than $1$ and decreases with
increasing $\delta$. When $\delta\rightarrow\infty$, the critical
value $\beta_c$ approaches a limit of
\begin{equation}
\beta_{cLim}\equiv\frac{m^2-2+(m^2+\alpha^2+1/4)} {m^2-2+{\cal
N}_m(\alpha)(m^2+\alpha^2+1/4)}\ .
\end{equation}
In parallel to the study of aligned cases, we further derive the
range of $\mu^g/\mu^s$ for both solution branches when $|m|$ and
$\delta$ are known at a given wavenumber $\alpha$. For the
$y_1\equiv D_s^2$ branch, we thus have
\begin{equation}
-1<\frac{\mu^g}{\mu^s}<\frac{-(m^2+\alpha^2+1/4) {\cal
N}_m(\alpha)\delta}{(m^2+\alpha^2+1/4) {\cal
N}_m(\alpha)\delta+(m^2-2)(1+\delta)}\qquad
\end{equation}
with the lower bound on the left-hand side determined by $\beta=1$
and the upper bound on the right-hand side determined by
$\beta\rightarrow\infty$; for the $y_2\equiv D_s^2$ solution
branch, we have
\begin{equation}
\delta<\frac{\mu^g}{\mu^s}<\frac{1+\delta}{{\cal
N}_m(\alpha)}-1\end{equation} with the lower bound on the
left-hand side determined by $\beta=1$ and the upper bound on the
right-hand side determined by the critical value $\beta_c$ such
that $y_2\equiv D_s^2=0$. Since $\mu^g/\mu^s$ increases with
increasing $\beta$, we always have negative $\mu^g/\mu^s$ for the
$y_1\equiv D_s^2$ branch and positive $\mu^g/\mu^s$ for the
$y_2\equiv D_s^2$ branch.

It is remarkable that the spiral cases are strikingly similar to
the aligned cases. If one replaces ${\cal N}_{m}(\alpha)$ by
$(m^2+\alpha^2+1/4)^{-1/2}$ and $(m^2+\alpha^2+1/4)^{1/2}$ by
$|m|$, all the above results for the spiral cases will degenerate
to the aligned cases. By these considerations of aligned and
spiral cases with $|m|\geq2$, this apparent similarity should not
be surprising because the common physical nature of stationary
aligned and spiral disturbances, that is, purely azimuthal and
general density waves in a composite SID system balanced by disk
rotation.

As promised earlier, we now get back to the $|m|=1$ case which is
special for a stationary spiral configuration as well as for a
stationary aligned configuration. For aligned $|m|=1$ case,
equation (51) is satisfied for arbitrary $D_s^2>0$ (see Shu et al.
2000 for a single SID). For spiral $|m|=1$ case, when $\delta$ and
$\alpha$ are given, the $y_1$ and $y_2$ solution branches will
increase and decrease with increasing $\beta$, respectively. When
$\beta\rightarrow\infty$, the two solutions are
$$
y=-1
$$
and
\begin{equation}
\quad y=-\bigg(\alpha^2+\frac{5}{4}\bigg) +\frac{{\cal
N}_1(\alpha)(\alpha^2+5/4)(\alpha^2+1/4)} {[{\cal
N}_1(\alpha)(\alpha^2+5/4)-1](1+\delta)}\ .\qquad
\end{equation}
Note in the above equation, we have ${\cal
N}_1(\alpha)(\alpha^2+5/4)-1$ always positive according to
equations (77) and (78).

The determination of the specific bounds of these two solutions
will depend on a relation between $\delta$ and $\alpha$. When
\begin{equation}
\begin{split}
y=-\bigg(\alpha^2+\frac{5}{4}\bigg) +\frac{{\cal
N}_1(\alpha)(\alpha^2+5/4)(\alpha^2+1/4)} {[{\cal
N}_1(\alpha)(\alpha^2+5/4)-1](1+\delta)}
\quad\qquad\qquad\nonumber\\
=-1-\bigg(\alpha^2+\frac{1}{4}\bigg) \frac{\delta[{\cal
N}_1(\alpha)(\alpha^2+5/4)-1]-1} {[{\cal
N}_1(\alpha)(\alpha^2+5/4)-1](1+\delta)}>-1 \qquad\nonumber
\end{split}
\end{equation}
which means $\delta<[{\cal N}_1(\alpha)(\alpha^2+5/4)-1]^{-1}$, we
have
\begin{equation}
-(\alpha^2+5/4)<y_1<-1\
\end{equation}
and
\begin{equation}
\begin{split}
-\bigg(\alpha^2+\frac{5}{4}\bigg) +\frac{{\cal
N}_1(\alpha)(\alpha^2+5/4)(\alpha^2+1/4)} {[{\cal
N}_1(\alpha)(\alpha^2+5/4)-1](1+\delta)}\\
<y_2<\frac{[1-{\cal N}_1(\alpha)](\alpha^2+5/4)}{{\cal
N}_1(\alpha)(\alpha^2+5/4)-1}\ .\qquad
\end{split}
\end{equation}
On the other hand, when
\begin{equation}
\begin{split}
y=-\bigg(\alpha^2+\frac{5}{4}\bigg) +\frac{{\cal
N}_1(\alpha)(\alpha^2+5/4)(\alpha^2+1/4)} {[{\cal
N}_1(\alpha)(\alpha^2+5/4)-1](1+\delta)}
\quad\qquad\qquad\nonumber\\
=-1-\bigg(\alpha^2+\frac{1}{4}\bigg) \frac{\delta[{\cal
N}_1(\alpha)(\alpha^2+5/4)-1]-1} {[{\cal
N}_1(\alpha)(\alpha^2+5/4)-1](1+\delta)}<-1 \qquad\nonumber
\end{split}
\end{equation}
which means $\delta>[{\cal N}_1(\alpha)(\alpha^2+5/4)-1]^{-1}$, we
have
\begin{equation}
\begin{split}
-\alpha^2-\frac{5}{4}<y_1<-\alpha^2-\frac{5}{4}+\frac{{\cal
N}_1(\alpha)(\alpha^2+5/4)(\alpha^2+1/4)} {[{\cal
N}_1(\alpha)(\alpha^2+5/4)-1](1+\delta)}\qquad\qquad
\end{split}
\end{equation}
and
\begin{equation}
-1<y_2<\frac{[1-{\cal N}_1(\alpha)](\alpha^2+5/4)}{{\cal
N}_1(\alpha)(\alpha^2+5/4)-1}\ .
\end{equation}
Note that for bounds of these inequalities, one side is determined
by $\beta=1$ and the other side by $\beta\rightarrow\infty$ for
both solution branches. It becomes clear that spiral $|m|=1$ case
differs from spiral $|m|\geq2$ cases. In particular, the ranges of
$y_1$ and $y_2$ solutions are controlled by a relation between
$\delta$ and $\alpha$, that is, whether $\delta>[{\cal
N}_1(\alpha)(\alpha^2+5/4)-1]^{-1}$ or not, such that the two
solution branches do not intersect with each other. By
inequalities (101) and (103),  the $y_1$ solution branch is always
negative and thus unphysical. In contrast, by inequalities (102)
and (104), it is still possible for portions of the $y_2$ solution
branch to be positive when $\delta$, $\alpha$ and $\beta$ take on
proper values. Note that this $y_2$ solution branch is the
counterpart of a single SID case.

As indicated in Figs. $11-13$, for given values of $|m|$, $\beta$
and $\delta$, it possible for a composite SID system to sustain
upper (smaller $\alpha$) and lower (larger $\alpha$) solution
branches with different values of $\alpha$ simultaneously.
Likewise, one may switch the role of $|m|$ and $\alpha$, that is,
given $\alpha$, a composite SID system can support various spiral
solutions with different values of $|m|$ (see Fig. 3 of Shu et al.
2000). This is to be understood that both $|m|$ and $\alpha$ are
characteristic properties of density waves rather than background
parameters of a composite SID system. In the linear regime, this
allowed superposition of different solutions can give rise to a
variety of galactic pattern structures that cannot fit by one
simple spiral pattern.

The interpretation for the stability properties of these
stationary logarithmic spiral configurations is not so
straightforward even for a single SID case because of the
power-law background divergence at $r\rightarrow 0$ and thus the
difficulty of imposing proper boundary conditions there (Zang
1976; Toomre 1977; Lynden-Bell \& Lemos 1999; Goodman \& Evans
1999; Shu et al. 2000; Galli et al. 2001). Shu et al. (2000)
speculated that the stationary spiral solution condition
represents onset of spiral instabilities on the ground that
transmission and overreflection of leading/trailing spiral density
waves across the corotation can be swing-amplified. This was done
mainly through numerical experiments and appeared to be consistent
with an earlier proposed criterion (Goodman \& Evans 1999).
Furthermore, the stability of stationary spiral configurations may
be examined in light of the Ostriker-Peebles criterion (Shu et al.
2000). It is then plausible to extend this interpretation to a
composite SID system where apparently more leading/trailing spiral
density waves across the corotation would be involved in the
processes of transmission and overreflection.

Therefore, in Figs. $11-13$, as $D_s^2$ increases from $0$ and
first intersects with the lower solution branch, the lower branch
spiral solution can be sustained and be vulnerable to dynamical
barred-spiral instabilities. As $D_s^2$ increases further to
intersect the upper solution branch, the upper branch spiral
solution (together with the lower branch solution at a larger
$\alpha$) can be sustained and be vulnerable to spiral
instabilities. In the case of Fig. 11, $D_s^2$ may be small enough
to avoid dynamical barred-spiral instabilities. But in cases of
Figs. 12 and 13, dynamical barred-spiral instabilities just cannot
be avoided for $D_s^2\geq 0$ when $\alpha$ is sufficiently large.


For the specific case of Fig. 11 with $m=2$, $\delta=0.2$ and
$\beta=1.5$, the two minima of the $y_1$ and $y_2$ solution
branches, i.e. the values of $D_s^2$ when $\alpha\rightarrow 0$,
are 1.3406 and 0.3697, respectively. The corresponding ratio of
${\cal T}/|{\cal W}|$ defined by equation (74) are 0.2982 and
0.1552 in order. In analogy to the single SID case of Shu et al.
(2000), the two minima of $D_s^2$ signal onsets of dynamical
barred-spiral instability in the composite SID system. By this
interpretation, when the rotation of the stellar SID $D_s^2$ is
smaller than 0.3697, the composite SID system can be stable
against $m=2$ spiral instabilities. Otherwise, dynamical
barred-spiral instabilities will develop in the composite SID
system, similar to processes shown in sections 5.7 and 7 of Shu et
al. (2000). Notice the introduction of the gaseous disk has
decreased somewhat the marginal value of $D_s^2$ and thus the
threshold ratio of ${\cal T}/|{\cal W}|$ in comparison to a single
SID case.

Similar to the aligned cases, spiral stability properties of a
composite SID system are more complicated than those of a single
SID. For stationary logarithmic spiral configurations, $D_s^2$
usually attains a minimum value at $\alpha=0$ for both upper and
lower solution branches of spiral equation (88), except for cases
when $y_2=D_s^2$ solutions become negative for small $\alpha$
values (see Figs. 5-9).

As in the aligned case, we start from the same parameter set of
$\delta=0.2$ and $\beta=1.5$ yet with a fixed $\alpha=0$. The
$y_2$ solution branch of spiral equation (88) then yields a curve
of $D_s^2$ versus $m\geq2$ as shown in Fig. 14 where the smallest
value of $D_s^2=y_2$ occurs at azimuthal periodicity $m=9$.

For the same $\delta=0.2$ and $\alpha=0$ yet a smaller
$\beta=1.2$, the curve of $D_s^2$ for $y_2$ solution branch of
spiral equation (88) versus $m\geq2$ is shown in Fig. 15, where
$m=2$ gives the smallest $D_s^2$.

\begin{figure}
\begin{center}
\includegraphics[angle=0,scale=0.29]{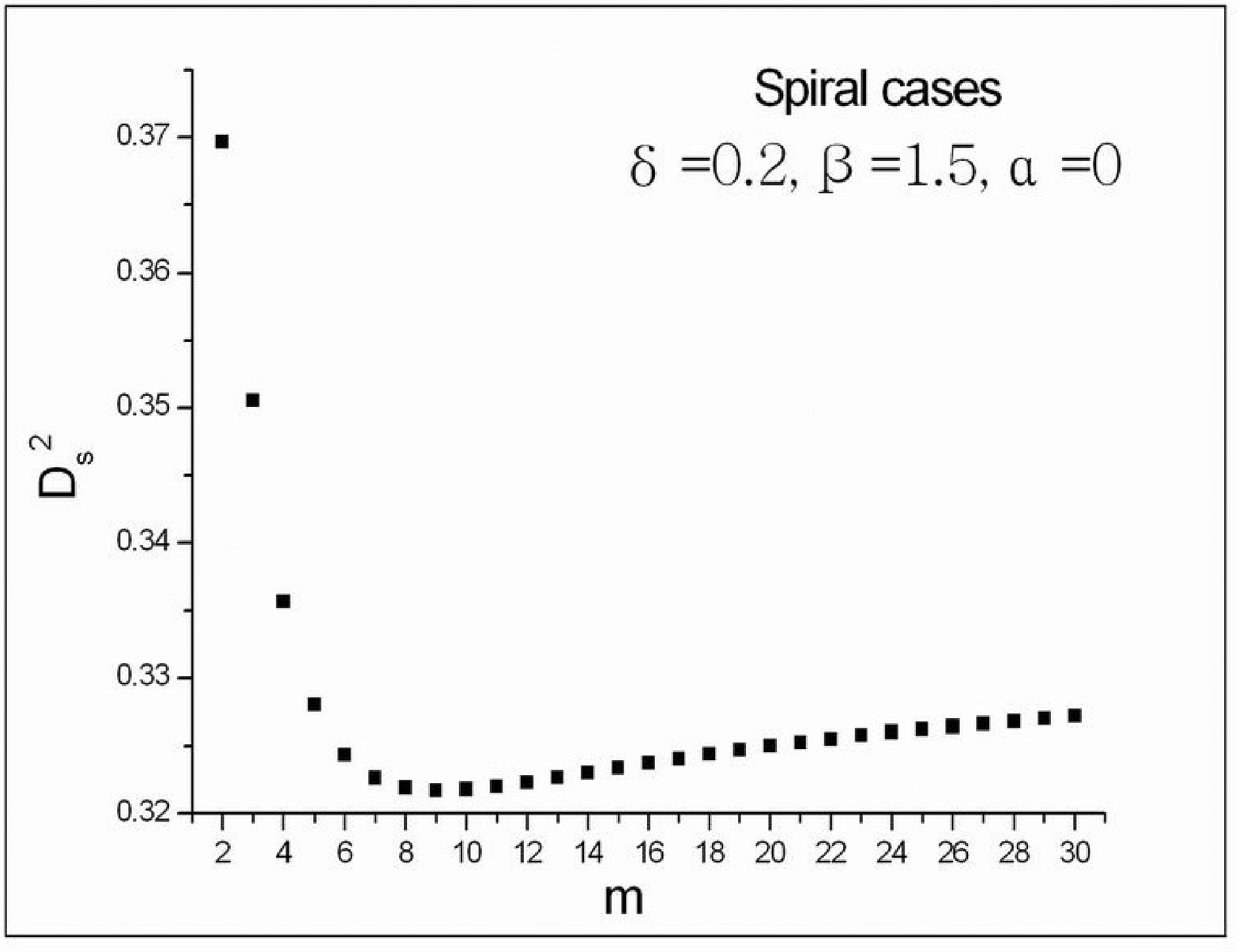}
\caption{The $D_s^2$ solution of $y_2$ branch from spiral equation
(88) versus azimuthal wavenumber $m\geq2$ with parameters
$\alpha=0$, $\delta=0.2$ and $\beta=1.5$. The smallest $D_s^2$ at
$m=9$.}
\end{center}
\end{figure}
\begin{figure}
\begin{center}
\includegraphics[angle=0,scale=0.29]{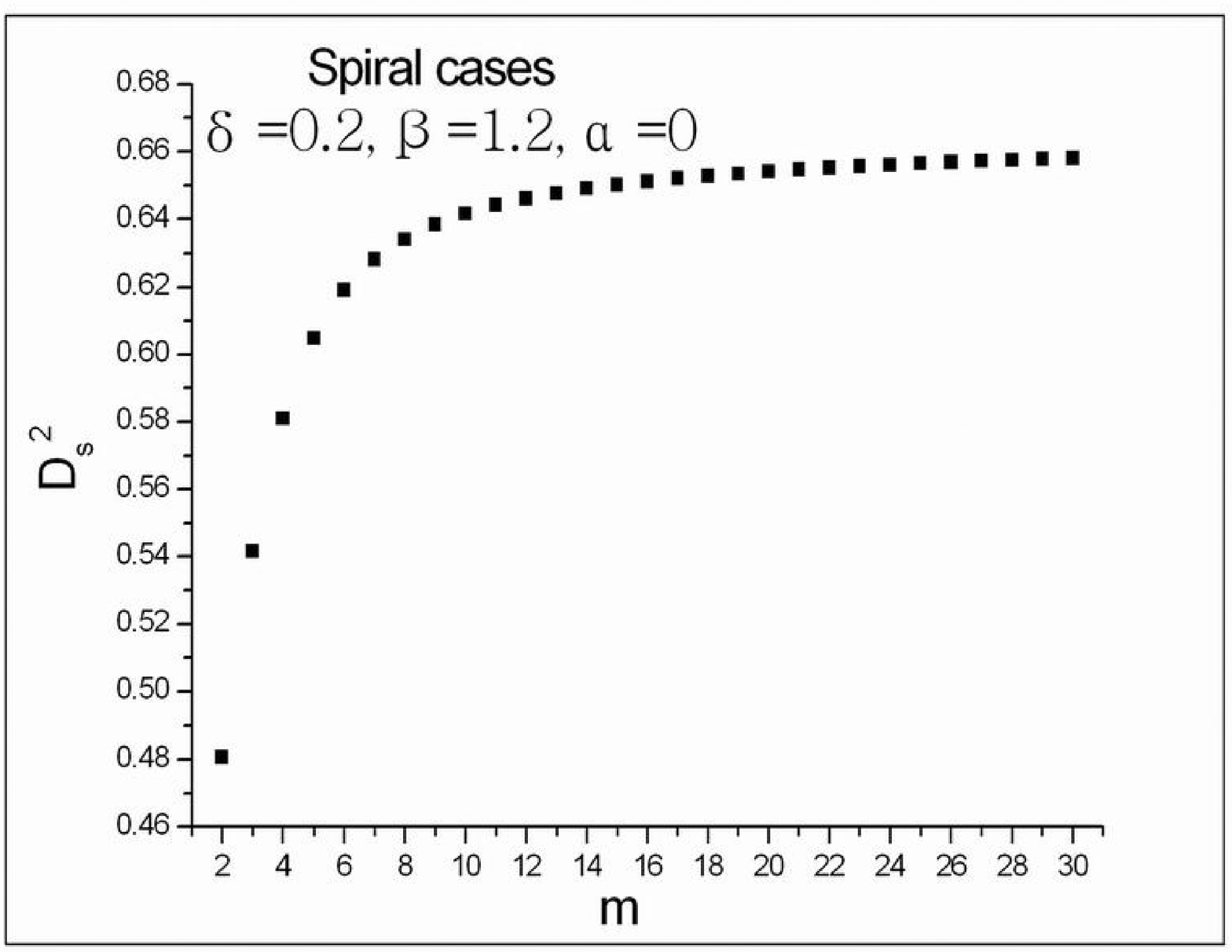}
\caption{The $D_s^2$ solution of $y_2$ branch from spiral equation
(88) versus $m\geq2$ with $\alpha=0$, $\delta=0.2$ and
$\beta=1.2$. The smallest $D_s^2$ at $m=2$.}
\end{center}
\end{figure}

\section{Composite partial SID system}

In this section, we consider stationary perturbations in a
composite system of two gravitationally coupled partial SIDs (Syer
\& Tremaine 1996; Shu et al. 2000; Lou 2002; Lou \& Fan 2002) to
model the gravitational influence of a background axisymmetric
dark-matter halo on the composite SID system. The response of this
massive dark-matter halo to perturbations in the composite SID
system is ignored. Formally, we introduce an additive gravity term
$\partial\Phi/\partial r$ or $\partial\Phi/\partial \varphi$ to
the basic equations (2), (3) and (5), (6), where $\Phi$ represents
the gravitational potential contribution from the background
dark-matter halo.

For momentum equations in the stellar disk, we have
\begin{equation}
\frac{\partial u^{s}}{\partial t}+ u^{s}\frac{\partial
u^{s}}{\partial r}+ \frac{j^{s}}{r^{2}}\frac{\partial
u^{s}}{\partial \varphi }-\frac{j^{s2}}
{r^{3}}=-\frac{1}{\Sigma^{s}} \frac{\partial }{\partial
r}(a_{s}^{2}\Sigma^{s})- \frac{\partial \phi }{\partial
r}-\frac{\partial \Phi} {\partial r}
\end{equation}
and
\begin{equation}
\frac{\partial j^{s}}{\partial t} +u^{s}\frac{\partial
j^{s}}{\partial r} +\frac{j^{s}}{r^{2}}\frac{\partial
j^{s}}{\partial\varphi} =-\frac{1}{\Sigma^{s}}\frac{\partial }
{\partial \varphi }(a_{s}^{2}\Sigma^{s}) -\frac{\partial \phi
}{\partial \varphi } -\frac{\partial \Phi}{\partial \varphi}\ .
\end{equation}
Likewise, for momentum equations in the gas disk, we have
\begin{equation}
\frac{\partial u^{g}}{\partial t} +u^{g}\frac{\partial
u^{g}}{\partial r} +\frac{j^{g}}{r^{2}}\frac{\partial
u^{g}}{\partial \varphi }
-\frac{j^{g2}}{r^{3}}=-\frac{1}{\Sigma^{g}} \frac{\partial
}{\partial r}(a_{g}^{2}\Sigma^{g}) -\frac{\partial \phi }{\partial
r} -\frac{\partial \Phi}{\partial r}
\end{equation}
and
\begin{equation}
\frac{\partial j^{g}}{\partial t} +u^{g}\frac{\partial
j^{g}}{\partial r} +\frac{j^{g}}{r^{2}}\frac{\partial
j^{g}}{\partial \varphi }
=-\frac{1}{\Sigma^{g}}\frac{\partial}{\partial\varphi}
(a_{g}^{2}\Sigma^{g})-\frac{\partial\phi}{\partial \varphi }
-\frac{\partial \Phi}{\partial \varphi}\ .
\end{equation}

For a composite partial SID system, we now introduce a
dimensionless ${\cal F}$ parameter defined by ${\cal F}\equiv
\phi/(\phi+\Phi)$. Now the background equilibrium should be
modified accordingly. As before, we write $\Omega_s =a_sD_s/r$,
$\Omega_g=a_gD_g/r$, and thus $\kappa_s=\sqrt{2}\Omega_s$,
$\kappa_g=\sqrt{2}\Omega_g$.
The equilibrium surface mass densities now become
\begin{equation}
\begin{split}
\Sigma_{0}^{s}=\frac{a_s^2}{2\pi Gr}
\frac{(1+D_s^2){\cal F}}{1+\delta}\ ,\\
\Sigma {}_{0}^{g}=\frac{a_{g}^{2}}{2\pi Gr}
\frac{(1+D_{g}^{2}){\cal F}\delta}{1+\delta}\ .
\end{split}
\end{equation}
Note that condition (11), namely, $a_s^2(D_s^2+1)=a_g^2(D_g^2+1)$
still holds. We refer to partial SIDs for $0<{\cal F}<1$, in
contrast to full SIDs with ${\cal F}=1$.

The linearized coplanar perturbation equations should take the
same forms of those in full SIDs written down in section 2.3. We
thus follow the procedure of the full SIDs case up to equations
(37) and (38). For the modified equilibrium with $\omega=0$, we
have for the stellar partial SID
\begin{equation}
\begin{split}
m\bigg[-\mu^s+\frac{1}{D_s^2(m^2-2)}
\bigg(\frac{m^{2}}{r}-2\frac{d}{dr}
-r\frac{d^{2}}{dr^{2}}\bigg)\qquad\\
\times\bigg(r\mu^{s}+\frac{1+D_s^2}{2\pi G} \frac{{\cal
F}V}{1+\delta}\bigg)\bigg]=0\ ,
\end{split}
\end{equation}
and for the gaseous SID
\begin{equation}
\begin{split}
m\bigg[-\mu^g +\frac{1}{D_g^2(m^2-2)}
\bigg(\frac{m^2}{r}-2\frac{d}{dr}
-r\frac{d^2}{dr^2}\bigg)\qquad\\
\times\bigg(r\mu^g+\frac{1+D_g^2}{2\pi G} \frac{\delta{\cal
F}V}{1+\delta}\bigg)\bigg]=0\
\end{split}
\end{equation}
(see equations (39) and (40)). Equations (110) and (111) are to be
solved simultaneously with Poisson's integral (32).

We now consider consequences of a composite partial SID system in
comparison with a composite system of two full SIDs.

For aligned cases: the $|m|=0$ disturbance is still a rescaling of
one axisymmetric equilibrium to a neighboring axisymmetric
equilibrium.
For aligned cases of $|m|=1$, the generalized counterpart of
equation (51) yields only one degenerate solution $y=-1$ which is
unphysical. In other words, eccentric displacements of $|m|=1$ are
not allowed in a composite partial SID system. With $\delta$ and
$|m|$ given, $|m|\geq2$ cases yield two solutions satisfying the
following inequalities
\begin{equation}
\begin{split}
\frac{m^2}{m^2-2}-\frac{2|m|(m^2-1){\cal F}}
{(m^2-2)(m^2+|m|{\cal F}-2)(1+\delta)}\\
<y_1< m^2/(m^2-2)
\end{split}
\end{equation}
and
\begin{equation}
-1<y_2<|m|(|m|-{\cal F})/(m^2+|m|{\cal F}-2)\ ,
\end{equation}
respectively (see inequalities (54) and (55) for the cases of full
SIDs). For a composite system of two coupled full SIDs with ${\cal
F}=1$, these results naturally reduce to the familiar ones,
namely, inequalities (54) and (55), derived earlier.

Given parameters $\delta$ and $\beta$ with $|m|\geq2$, partial
SIDs lead to two solution branches of $D_s^2$ that are larger than
those of full SIDs. Therefore, the critical value $\beta_c$ for
$y_2\equiv D_s^2=0$ is larger in partial SIDs, namely
\begin{equation}
\begin{split}
\!\!\!\! \beta_c=2(m^2-1)/(m^2+|m|{\cal F}-2)
\qquad\qquad\qquad\qquad\qquad\\
\times\bigg[1+\frac{|m|(|m|-1){\cal F}} {|m|(m^2+|m|{\cal
F}-2)\delta +(m^2-2)(|m|-{\cal F})}\bigg]\ .\qquad
\end{split}
\end{equation}
For full SIDs with ${\cal F}=1$, expression (114) reduces to
expression (56) as expected. Meanwhile, the ratio of surface mass
density perturbations $\mu^g/\mu^s$ becomes
\begin{equation}
\frac{\mu^g}{\mu^s}
=-1-\frac{[D_s^2(m^2-2)-m^2](1+\delta)}{|m|(D_s^2+1){\cal F}}\ .
\end{equation}
Specifically for the two solution branches of $D_s^2$, we derive
two inequalities, namely
\begin{equation}
-1<\frac{\mu^g}{\mu^s}<-\frac{|m|{\cal F}\delta} {(m^2+|m|{\cal
F}-2)\delta+m^2-2}
\end{equation}
for the $D_s^2=y_1$ solution branch, and
\begin{equation}
\delta<\frac{\mu^g}{\mu^s}<|m|(1+\delta)/{\cal F}-1
\end{equation}
for the $D_s^2=y_2$ branch. Notice that expressions $(115)$,
$(116)$ and $(117)$ have their obvious counterparts of the full
SIDs cases of expressions $(53)$, $(58)$ and $(59)$ derived
earlier.

The same procedure can be applied to the unaligned situation,
which is expected to yield the same trends as the aligned
situation.

We now derive the virial theorem in a composite partial SID
system. By equations (105) and (107) for the equilibrium state, we
obtain
\begin{equation}
-\Sigma_0^sr\Omega_s^2=-\frac{d}{dr}(a_s^2\Sigma_0^s)
-\Sigma_0^s\frac{d(\phi_0+\Phi)}{dr}
\end{equation}
and
\begin{equation}
-\Sigma_0^gr\Omega_g^2=-\frac{d}{dr}(a_g^2\Sigma_0^g)
-\Sigma_0^g\frac{d(\phi_0+\Phi)}{dr}\ .
\end{equation}
Adding equations $(118)$ and $(119)$, multiplying the resulting
equation by $2\pi r^2dr$ and integrating from $0$ to $R$, we
obtain
\begin{equation}
2({\cal T}+{\cal U})+{\cal W}=2\pi
R^2[a_s^2\Sigma_0^s(R)+a_g^2\Sigma_0^g(R)]\ ,
\end{equation}
where
\begin{equation}
{\cal T}\equiv\int_0^R\frac{1}{2}\Sigma_0^s(r\Omega_s)^22\pi
rdr+\int_0^R\frac{1}{2}\Sigma_0^g(r\Omega_g)^22\pi rdr\ ,
\end{equation}
\begin{equation}
{\cal U}\equiv\int_0^R(a_s^2\Sigma_0^s+a_g^2\Sigma_0^g)2\pi rdr\ ,
\end{equation}
\begin{equation}
{\cal W}\equiv-\int_0^Rr(\Sigma_0^s+\Sigma_0^g)
\frac{d(\phi_0+\Phi)}{dr}2\pi rdr
\end{equation}
are the kinetic energy of rotation, the thermal energy and the
gravitational potential energy, respectively. Note here the
dark-matter halo contributes to the gravitational potential energy
and ${\cal F}\equiv \phi_{0}/(\phi_{0}+\Phi)$.

%

\section{Summary and discussions}

There are four main tracks of research that lead to our
investigation here on stationary perturbation configurations in a
composite system of two gravitationally coupled full or partial
SIDs. The first track is the classical study on spheroidal and
ellipsoidal figures of equilibrium (Chandrasekhar 1969; Tassoul
1978) named after famous mathematicians including Maclaurin,
Jacobi, Dedekind, Riemann and Poincar\'e. These spheroids and
ellipsoids have their analogous compressible counterparts
(Ostriker 1978), disk counterparts by collapsing along the
rotation axis (Weinberg \& Tremaine 1983; Weinberg 1983), as well
as counterparts of composite systems (Vandervoort 1991a,b). The
second track is along the past investigation of possible SID
configurations and their stability properties (Zang 1976; Toomre
1977; Lemos, Kalnajs \& Lynden-Bell 1991; Lynden-Bell \& Lemos
1993; Lee \& Goodman 1999; Evans \& Read 1999; Goodman \& Evans
1999; Syer \& Tremaine 1996; Shu et al. 2000; Galli et al. 2001).
The third track follows the studies of density waves in a
composite disk system (Lin \& Shu 1966; Jog \& Solomon 1984a,b;
Bertin \& Romeo 1988; Elmegreen 1995; Jog 1996; Lou \& Fan 1998b,
2000a,b). The fourth track pursues a basic understanding of MHD
density waves in magnetized disk systems (Fan \& Lou 1996; Lou \&
Fan 1998a, 2002, 2003; Shu et al. 2000; Lou, Yuan \& Fan 2001; Lou
2002). Naturally, we shall further study possible stationary
configurations in a composite system of gravitationally coupled
stellar SID with gaseous MSID. These studies are of interests in
galactic dynamics (Bertin \& Lin 1996), star formation (Shu et al.
1999), the light cusps seen in the nuclei of galaxies (Crane et
al. 1993), circumnuclear starburst ``rings" and disk accretion
processes around active galactic nuclei (Lou et al. 2001).

In reference to the work of Lou \& Fan (1998b) on density waves in
a composite disk system, of Shu et al. (2000) on stationary
perturbation configurations of a single isopedically magnetized
SID, and of Lou (2002) on stationary logarithmic fast and slow
configurations in a single MSID, we studied here possible
stationary configurations in a composite system of two
gravitationally coupled full or partial SIDs using the two-fluid
formalism, solved the Poisson equation exactly, and derived both
aligned and unaligned or spiral stationary perturbation solutions.
We have reached several conclusions and suggestions summarized
below.

\begin{enumerate}
\item Aligned Stationary Configurations

For perturbation configurations in a composite SID system, there
exist in general two branches of solutions as expected (Lou \& Fan
1998b). This is one major difference between a composite SID
system and a single SID case. We shall start with the aligned
cases first.

The aligned axisymmetric $|m|=0$ case is again a rescaling from
one axisymmetric equilibrium to a neighboring axisymmetric
equilibrium (Shu et al. 2000; Lou 2002), except that now the
rescaling occurs simultaneously in both SIDs.

We find as in the single full SID case (Shu et al. 2000) that, for
eccentric $|m|=1$ displacements, equation (51) can be satisfied
for unconstrained positive values of $D_s^2$. Given known
idealizations of the full SID model, we consider that such
eccentric displacements are possible. However, the situation is
quite different for a composite system of two partial SIDs for
which eccentric displacements are not allowed consistent with the
similar result of a single partial SID (Lou 2002).

For $|m|\geq2$, we derive two branches of $D_s^2$ solution for
possible values of rotation speed parameter $D_s$ such that
aligned stationary configurations are sustained in a composite SID
system. Of the two branches, one is always larger than the other.
For the larger branch denoted by $y_1$, perturbation enhancements
of surface mass densities in stellar and gaseous disks are out of
phase, while for the smaller branch denoted by $y_2$, the two
enhancements are in phase. By the physical requirement of
$D_s^2>0$, the larger $y_1$ branch is always valid (see inequality
54), while the smaller $y_2$ branch may become negative and thus
unphysical (see inequality 55). Given $|m|\geq2$ and $\delta$,
when $\beta$ exceeds a critical value $\beta_c$ defined by
equation (56), the smaller $y_2$ branch turns negative (see Fig.
1). In other words, only for disk galaxies with $\beta<\beta_c$,
the smaller $y_2$ branch can be physically meaningful. Moreover,
the smaller $y_2$ branch corresponds to the solution of single SID
case as it would approach the limit of the single SID result when
$\beta\rightarrow 1$, that is, when the system can be effectively
viewed as a single SID. In contrast, the larger $y_1$ branch
obtained in a composite SID system has no counterpart in the
single SID case. Physically, stationary aligned disturbances
actually involve purely azimuthal propagation of density waves
counter balanced by SID rotation (Lou 2002; Lou \& Fan 2002). The
same density-wave interpretation holds for stationary unaligned or
spiral perturbations. As in the single SID case, stationary
aligned configurations here mark onsets of secular instabilities
from an axisymmetric equilibrium to nonaxisymmetric
configurations.
For different values of $|m|$, distinct aligned configurations may
coexist.

\item Unaligned or Spiral Stationary Configurations

Solution properties of the unaligned or spiral logarithmic
configurations bear strong resemblance to those of the aligned
cases. The spiral case involves both radial and azimuthal density
wave propagations, whereas the aligned case involves only an
azimuthal propagation of density waves. To a certain extent, we
may regard aligned perturbation cases as special examples of
spiral perturbation cases with $|m|\geq2$.

For the marginal case of $|m|=0$, we have computed the marginal
stability curves of $D_s^2$ versus radial wavenumber $\alpha$,
invoking the same physical interpretations of Shu et al. (2000)
for regimes of collapse, stable oscillations, and ring
fragmentation. We shall further demonstrate the correctness of
this interpretation by performing a WKBJ analysis in a separate
paper. For different parameters $\delta$ and $\beta$, we find
these marginal stability curves may vary significantly in
reference to those obtained by Shu et al. (2000). Numerical
examples are illustrated in Figs. $5-10$ (see also Fig $D1$ in
Appendix D for a typical solution structure). Generally speaking,
the introduction of a gaseous SID tends to decrease the stability
of the composite SID system. Meanwhile, the possibility for the
system to collapse may be suppressed (Fig. 9). In primordial disk
galaxies where gas disk mass exceeds stellar disk mass, the
possibility of large-scale collapse may be suppressed while the
chance of ring fragmentation increases.

The $|m|=1$ case is again similar to the single SID problem.
Specifically, the $y_1$ solution branch, which is the novel
solution of a two-SID system, remains always negative and is thus
unphysical. Meanwhile the $y_2$ solution branch, which is the
counterpart of one-SID problem, can still have physically valid
portions constrained by inequalities (102) and (104).

For $|m|\geq2$ cases, we have obtained analytical results almost
in the same forms of aligned cases. This can be clearly seen by
comparing equations $(94)-(99)$ with equations $(54)-(59)$. By
replacing the Kalnajs function ${\cal N}_m(\alpha)$ with
$(m^2+\alpha^2+1/4)^{-1/2}$ and $(m^2+\alpha^2+1/4)^{1/2}$ with
$|m|$ in results of spiral cases, the resulting equations become
the same as the aligned cases. To reiterate, the aligned case is
just a special one complementary to spiral cases, all of which
involve hydrodynamic density waves in a composite system of two
coupled SIDs.

It should be mentioned that in the WKBJ regime, such stationary
logarithmic spiral density wave perturbations actually carry an
outgoing angular momentum flux which is constant in $r$. With our
composite SID model, a source of angular momentum is thus required
at the origin $r=0$ where the surface mass density diverges as
$\sim r^{-1}$. More generally, the outward advective angular
momentum flux (stellar and gas SIDs together) is constant in $r$,
while the gravity torque term involves a nontrivial integral in
$z$ (see Section 4 of Fan \& Lou 1999). In various contexts,
several authors, e.g. Lynden-Bell \& Kalnajs (1972), Goldreich \&
Tremaine (1978), and Fan \& Lou (1999), have previously
investigated angular momentum fluxes carried by density waves or
MHD density waves in the tight-winding or WKBJ regime.

\item A Composite System of Two Partial SIDs

In contexts of disk galaxies, the gravitational effect of the more
massive dark-matter halo may be incorporated using a model of
so-called partial SID by introducing a ${\cal F}$ parameter which
is the ratio of disk potential to the total potential ($0<{\cal
F}\leq 1$). The background surface mass densities of the two
coupled SIDs are modified. For stationary perturbations in a
composite partial SID system ($0<{\cal F}< 1$), it is
straightforward to make use of the full SID formalism (${\cal
F}=1$) by replacing $1/(1+\delta)$ with ${\cal F}/(1+\delta)$ and
$\delta/(1+\delta)$ with ${\cal F}\delta/(1+\delta)$ in equations
(39) and (40).

Naturally, all the previous full SID results are modified by the
${\cal F}$ parameter quantitatively. In particular, the aligned
$|m|=1$ case is no longer a solution for arbitrary positive
$D_s^2$, that is, such stationary eccentric displacements are
forbidden in a composite system of two partial SIDs. Also, the
critical value $\beta_c$ for physically valid $y_2$ solution
branch (both aligned or unaligned) with $|m|\geq2$ is increased
somewhat.

While our model formulation is highly idealized in several
aspects, it is one step closer to a realistic disk galaxy. The
theoretical results here provide a few useful concepts and
possible clues for diagnostics of disk galaxies composed of stars
and gas coupled by mutual gravity on large scales. Analytical
solutions derived here offer valuable benchmarks for numerical
codes designed for N-body computations with fluid gas dynamics.
The analysis here also paves the way to study theoretical model
problems of stationary configurations in a composite system of a
stellar SID gravitationally coupled with a gaseous MSID (Lou
2002). It is of considerable interest to extend the current
analysis into the nonlinear realm (Galli et al. 2001) to explore
the large-scale dynamics of spiral shocks.

\item Stability of a Composite SID System

For both aligned and spiral instabilities, a composite SID system
is more complicated than a single SID system in several aspects.
Firstly, roughly speaking, the number of possible stationary
solutions is doubled except for unphysical $y_2$ solutions that
must be excluded. Secondly, there are two more dimensionless
parameters $\beta$ and $\delta$ that may cause variations in
solution structures. Thirdly, because a composite SID system tends
to be less stable, $D_s^2$ solutions of lower $y_2$ branches may
correspond to ${\cal T}/|{\cal W}|$ ratios considerably less than
$\sim 0.14$ (Ostriker \& Peebles 1973). Fourthly, in certain
parameter regimes of $\delta$ and $\beta$ with $m=0$, the collapse
regime of spiral cases (Shu et al. 2000) may disappear. We shall
further examine in detail axisymmetric stability properties of a
composite SID system in a separate paper. Fifthly, because of
variations in solution structures, one needs to be very careful to
determine the most vulnerable configurations with smallest $D_s^2$
values. We propose to test the Ostriker-Peebles criterion for a
composite disk system by numerical simulation experiments and
expect considerably lower ${\cal T}/|{\cal W}|$ ratio than $\sim
0.14$ in certain parameter regimes studied here.

\item  Dynamics of Magnetized SIDs

For the evolution of an individual spiral galaxy, star formation
is one major process (e.g., Kennicutt 1989) and the global star
formation rate is a key parameter. In the density wave scenario
(Bertin \& Lin 1996), spiral arms of higher gas concentration are
locations of active star forming regions on smaller scales.
Therefore, a protodisk galaxy begins with a gas disk will
eventually evolve into a stellar disk. In between the two
extremes, the disk galaxy involves a gas disk and a stellar disk
with a variable mass ratio. As time goes on, their mass ratio as
well as the ratio of velocity dispersions change. At various
stages, large-scale instabilities may occur in a composite disk
system to alter the evolution track. Furthermore, the presence of
magnetic field in the gas disk may give rise to
magnetohydrodynamic (MHD) density waves (Fan \& Lou 1996; Lou \&
Fan 1998) and instabilities that can significantly affect the star
formation rate.

Magnetic field effects are not included at present merely for
simplicity. In disk galaxies, the gaseous disk is magnetized with
energy densities of magnetic field, thermal gas and cosmic rays
being comparable (Lou \& Fan 2003). We derive such information
based on synchrotron radio emissions from spiral galaxies.
Large-scale magnetic fields typically lie in the disk plane, but
there are cases with out-of-plane magnetic fields. On various
scales, MHD will play an important role in the gaseous disk and
provide valuable dynamical as well as diagnostic information (Fan
\& Lou 1996; Lou \& Fan 1998a; Lou et al. 2001, 2002; Lou 2002).
In the current contexts, we simply point out two possible
extensions of this theoretical analysis. One is to consider
stationary perturbation configurations in a composite partial SID
system in which the gas disk is isopedically magnetized (Shu et
al. 2000). It is then possible to study the influence of magnetic
field on the stability of the system. Such a problem is
potentially important to the development of galactic wind models.
The other is to consider stationary perturbation configurations in
a composite partial SID system in which the gas disk is magnetized
with coplanar magnetic fields (Lou 2002; Lou \& Fan 2002). The
presence of magnetic fields allows for more dynamic freedom in the
gas disk and gives rise to fast and slow MHD density waves (Fan \&
Lou 1996; Lou et al. 2001). We shall pursue these more realistic
yet challenging MHD model problems in separate papers.

\end{enumerate}

\section*{Acknowledgments}

This research has been supported in part by the ASCI Center for
Astrophysical Thermonuclear Flashes at the University of Chicago
under Department of Energy contract B341495, by the Special Funds
for Major State Basic Science Research Projects of China, by the
Tsinghua Center for Astrophysics, by the Collaborative Research
Fund from the NSF
of China (NSFC) for Young Outstanding Overseas Chinese Scholars
(NSFC 10028306) at the National Astronomical Observatory, Chinese
Academy of Sciences, and by the Yangtze Endowment from the
Ministry of Education through the Tsinghua University. Affiliated
institutions of Y.Q.L. share the contribution.

\appendix
\section\\

For the convenience of reference, we here outline a general proof
that equation (51) for stationary aligned perturbations always
have two branches of real solutions of $y\equiv D_s^2$ when
$|m|\geq 2$. Note that $m=0$ is a rescaling of the background and
$m=1$ leads to unconstrained $D_s^2$. By multiplying a factor
$(1+\delta)$, the coefficients of quadratic equation (51) can be
rewritten in the following forms:
\begin{equation}
C_2'=\beta(m^2-2)(m-1)(m+2)(1+\delta)\ ,
\end{equation}
\begin{equation}
\begin{split}
C_1'=&-2(m-1)[(m+2)\delta-(m^2-2)]\beta\\
&-2(m^2-1)[(m^2-2)\delta+m^2+m-2]\ ,
\end{split}
\end{equation}
\begin{equation}
\begin{split}
C_0'=-m(m&-1)[m(m+2)\delta+m^2-2]\beta\\
&+2m(m^2-1)(m\delta+m-1)\ ,
\end{split}
\end{equation}
where the common factor $(m-1)$ is made explicit in all three
coefficients. The determinant $\Delta$ of equation (51) is then
given by
\begin{eqnarray}
\qquad\Delta&\equiv&C_1^{'2}-4C_2'C_0'\nonumber\\
\qquad&=&4(m^2-1)^2({\cal A}\beta^2+{\cal B}\beta+{\cal C})\ ,
\end{eqnarray}
where three coefficients ${\cal A}$, ${\cal B}$, and ${\cal C}$
are defined by
\begin{equation}
{\cal A}\equiv [m^2-2+(m^2+m-2)\delta]^2\ ,
\end{equation}
\begin{equation}
\begin{split}
{\cal B}=&-2(m^2-2)(m^2+m-2)(\delta^2+1)
\\&
-2(2m^4+2m^3-9m^2-4m+8)\delta\ ,
\end{split}
\end{equation}
\begin{equation}
{\cal C}=[m^2+m-2+(m^2-2)\delta]^2\ .
\end{equation}
We can now rewrite equation (A4) in the form of
\begin{equation}
\Delta=4(m^2-1)^2 \bigg[{\cal A}\bigg(\beta+\frac{{\cal B}}{2{\cal
A}}\bigg)^2 +{\cal C}-\frac{{\cal B}^2}{4{\cal A}}\bigg]\ .
\end{equation}
Since both ${\cal A}>0$ and
\begin{equation}
{\cal C}-\frac{{\cal B}^2}{4{\cal A}}
=\frac{4m^2(m^2-2)(m^2+m-2)\delta(1+\delta)^2}
{[m^2-2+(m^2+m-2)\delta]^2}>0
\end{equation}
for $m\geq 2$, the determinant of equation (51) is clearly always
positive. Therefore, equation (51) will give two branches of real
solutions for $D_s^2=y$ for $m\geq 2$.

In the same spirit, we can show numerically that for unaligned or
spiral cases of $|m|\geq1$, equation (88) always has two branches
of real solutions for $D_s^2=y$. To avoid repetition and
redundancy, we omit that demonstration here.


\section\\

We demonstrate here that when $|m|\geq 2$, the two branches of
solution of equations (51) and (88) monotonically decrease with
increasing $\beta$ as shown in Fig. 1.

Let us consider equation (51) for aligned cases with fixed $|m|$
to simplify the computation. For example, with $|m|=2$ in equation
(51), the three coefficients of equation (51) as modified in
Appendix A become
\begin{equation}
C''_2=2(1+\delta)\beta\ ,
\end{equation}
\begin{equation}
C''_1=-(2\delta-1)\beta-3(\delta+2)\ ,
\end{equation}
\begin{equation}
C''_0=-(4\delta+1)\beta+3(2\delta+1)
\end{equation}
by further removing a common factor 4. Explicitly, the two
solutions of equation (51) can then be written as
\begin{equation}
y_{1,2}=\frac{-C''_1\pm (C^{''2}_1-4C''_2C''_0)^{1/2}}{2C''_2}\ .
\end{equation}
Substituting expressions $(B1)-(B3)$ for $C''_2$, $C''_1$ and
$C''_0$ into equation $(B4)$ and taking the first partial
derivative with respect to  $\beta$, we find that for any positive
$\delta$, $\partial y_1/\partial\beta$ and $\partial
y_2/\partial\beta$ are always negative. Therefore, both $y_1$ and
$y_2$ solution branches decrease with increasing $\beta$. The
$|m|\geq3$ cases can be proven in a similar manner.

For spiral cases, we follow the same procedure to find that for
$|m|\geq2$ cases both $y_1$ and $y_2$ solution branches decrease
with increasing $\beta$ while for the special case of $|m|=1$, the
$y_1$ and $y_2$ solution branches increase and decrease with
increasing $\beta$, respectively.

\section\\

One can readily show that $\mu^g/\mu^s$ decreases with increasing
$D_s^2$ for both aligned and unaligned cases when $|m|\geq2$ and
$|m|\geq 1$, respectively. For aligned cases, we simply rearrange
equation (53) into the following form of
\begin{equation}
\frac{\mu^g}{\mu^s}=-1-\frac{(m^2-2)(1+\delta)}{|m|}
+\frac{2(m^2-1)(1+\delta)}{|m|(D_s^2+1)}
\end{equation}
which shows obviously that $\mu^g/\mu^s$ decreases with increasing
$D_s^2$ when $|m|\geq2$.

For spiral cases, we can also arrange equation (92) into the
parallel form of
\begin{equation}
\begin{split}
\frac{\mu^g}{\mu^s}=&-1-\frac{(m^2-2)(1+\delta)}{{\cal
N}_m(\alpha)(m^2+\alpha^2+1/4)}\\
&+\frac{[m^2-2+(m^2+\alpha^2+1/4)](1+\delta)} {{\cal
N}_m(\alpha)(m^2+\alpha^2+1/4)(D_s^2+1)}
\end{split}
\end{equation}
which again shows clearly that $\mu^g/\mu^s$ decreases with
increasing $D_s^2$ when $|m|\geq 1$ for any radial wavenumber
$\alpha$. Note in particular that when $|m|=1$, the coefficient of
$1/(D_s^2+1)$ on the right-hand side of equation $(C2)$ is
positive.

\section\\

In general, for two similar dynamical systems coupled together, we
expect two sets of oscillation modes (e.g., Vandervoort 1991a, b).
This should be true for the current problem of a composite SID
system as compared to a single SID system. For the cases of
marginal stability for spiral configurations with $m=0$, we
thought that there might be something interesting from the
additional solution of a composite SID system in contrast to the
single SID case (Shu et al. 2000). It turns out that this
additional solution of $D_s^2$ is negative by numerical
exploration.
The numerical example of Figure $D1$ shows the typical structure
of solution curves for stationary logarithmic spiral
configurations when $m=0$, $\delta=0.2$ and $\beta=1.5$. For
mathematical completeness, the negative solution branches are
shown explicitly. In Fig. 5 of the main text, the negative
solutions of $D_s^2$ are discarded.
\begin{figure}
\begin{center}
\includegraphics[scale=0.35]{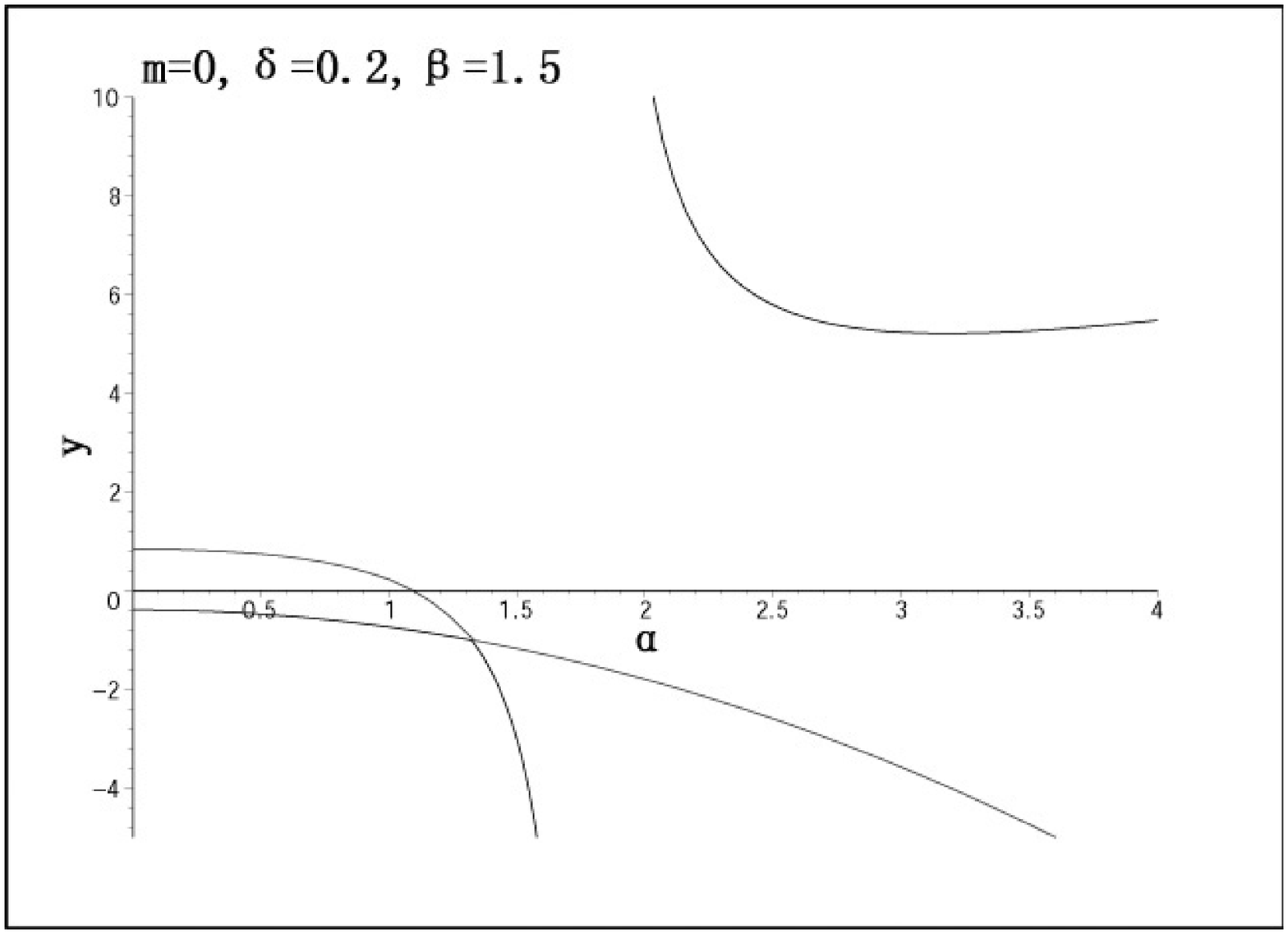}
\caption{The complete solution curves of marginal stability for
spiral cases in a composite disk system when $m=0$, $\delta=0.2$
and $\beta=1.5$. The negative branches are shown here explicitly
on purpose.}
\end{center}
\end{figure}

\end{document}